\documentclass[11pt,draftcls,onecolumn]{IEEEtran}
\usepackage{booktabs}
\usepackage{amsfonts,bm,amsmath,comment,color,cite,amssymb,subfigure,times}
\usepackage[]{graphicx}
\usepackage[linesnumbered,ruled]{algorithm2e}
\usepackage{multirow}

\def\ba{{\mathbf a}}
\def\bb{{\mathbf b}}

\def\bd{{\mathbf d}}

\def\bf{{\mathbf f}}

\def\bh{{\mathbf h}}

\def\bn{{\mathbf n}}

\def\bp{{\mathbf p}}
\def\bq{{\mathbf q}}
\def\br{{\mathbf r}}
\def\bs{{\mathbf s}}

\def\bw{{\mathbf w}}
\def\bx{{\mathbf x}}
\def\by{{\mathbf y}}

\def\bA{{\mathbf A}}
\def\bB{{\mathbf B}}

\def\bH{{\mathbf H}}
\def\bI{{\mathbf I}}

\def\bR{{\mathbf R}}
\def\bS{{\mathbf S}}
\def\bT{{\mathbf T}}

\def\bX{{\mathbf X}}
\def\bY{{\mathbf Y}}
\def\bZ{{\mathbf Z}}
\def\ups{\boldsymbol{\upsilon}}
\def\bnu{\boldsymbol{\nu}}
\def\complexC{{\mathbb{C}}}
\def\realR{{\mathbb{R}}}
\def\expE{{\mathbb{E}}}

\def\tbx{{{\widetilde \bx}}}

\def\tbT{{{\widetilde \bT}}}
\def\hbx{{{\hat{\bx}}}}

\def\bzero{{\mathbf{0}}}

\begin{document}

\title{Constant-Modulus Waveform Design for Dual-Function Radar-Communication Systems in the Presence of Clutter}
\author{Wenjun Wu, Bo Tang, Xuyang Wang
\thanks{This work was supported in part by the National Natural Science Foundation of China under Grant 62171450 and 61671453, the Anhui Provincial Natural Science Foundation under Grant 2108085J30, the Young Elite Scientist Sponsorship Program of CAST under Grant 17-JCJQ-QT-041, and the Postgraduate Scientific Research Innovation Project of Hunan Province under Grant CX20220073 (\it{Corresponding author: Bo Tang}\rm).}
\thanks{ Wenjun Wu, Bo Tang and Xuyang Wang are with the College of Electronic Engineering, National University of Defense Technology, Hefei 230037, China. Email: tangbo06@gmail.com.}
\thanks{}}

\markboth{Preprint}
{Shell \MakeLowercase{\textit{et al.}}: A Sample Article Using IEEEtran.cls for IEEE Journals}

\maketitle

\begin{abstract}
We investigate the constant-modulus (CM) waveform design for dual-function radar communication systems in the presence of clutter.
To minimize the interference power and enhance the target acquisition performance, we use the signal-to-interference-plus-noise-ratio as the design metric.
In addition, to ensure the quality of the service for each communication user, we enforce a constraint on the synthesis error of every communication signals.
An iterative algorithm, which is based on cyclic optimization, Dinkinbach's transform, and alternating direction of method of multipliers, is proposed to tackle the encountered non-convex optimization problem.
Simulations illustrate that the CM waveforms synthesized by the proposed algorithm allow to suppress the clutter efficiently and control the synthesis error of communication signals to a low level.
\end{abstract}


\section{Introduction}

With the rapid development of 5G and 6G networks, autonomous driving, and internet of things (IOT), we have witnessed the widely development of active sensing systems (such as radar and communication systems).
Typically, active sensing systems utilize the radio frequency (RF) spectrum to determine the property of the target or the propagation medium.
However, because of the scarceness of the RF spectrum, the increasing number of active sensing systems makes the spectrum extremely crowded. For active systems operating in spectrally crowded environments, the mutual interference will severely degrade the system performance \cite{Griffiths2014spectrum,Griffiths2015spectrum}.
Therefore, improving the coexistence between active sensing systems has gained considerable interests in the past few years (see, e.g., \cite{Aubry2016Optimization,Qian2018Coexistence,Zheng2019Coexistence} and the references therein).

Many paradigms have been implemented to improve the spectral coexistence between radar and communication systems.
One possibility is to use the opportunistic illuminator (e.g., signals from satellites, broadcast transmitters, base stations, etc.) for radar detection.
Since such radar systems (also called passive radar) receive the direct-path signals and the target reflections passively to detect the targets \cite{howland2005passive,kuschel2019tutorial}, the conflict between radar and communication systems could be minimized.
However, passive radar systems might suffer the problem of low range resolutions and high sidelobes.
Recently, cognitive paradigms are extensively discussed to improve the spectral compatibility among active sensing systems \cite{Jakabosky2016spectrum,Aubry2021Cognitive}.
Through the perception of outside environments, cognitive systems can minimize the mutual interference, e.g., by adaptive processing in the receiver, or adjusting the waveforms intelligently in the transmitter \cite{Shi2022TAES}.
In \cite{Lindenfeld2004Sparse,Rowe2014SHAPE,Aubry2014spectrally,Liang2015LPNN,Tang2019Efficient}, by assuming that the operating frequency bands of nearby communication systems can be obtained through spectrum sensing, methods for synthesizing spectrally constrained waveforms were proposed.
The proposed waveforms can form deep notches in the stopbands (i.e., the operating frequency bands of the communication systems), thus improving the spectral coexistence between radar and nearby communication systems.
In \cite{Li2016codesign,Li2017Coexistence,Qian2018Coexistence}, the co-design of multiple-input-multiple-output (MIMO) radar and MIMO communication systems were discussed.
By sensing the channel state as well as the clutter/interference parameters, the co-design paradigm can improve the target acquisition performance and ensure the quality of service (QOS) for communications.

In addition to the aforementioned paradigms, there are many attempts in pursuing the functional coexistence.
Herein, the functional coexistence refers to using an integrated system to simultaneously support multiple functions, including radar and communications.
Such systems are also called dual-function radar-communication (DFRC) systems, or integrated sensing and communication (ISAC) systems \cite{Tavik2005RF, Zhang2022Survey,Zhang2021Overview,Liu2020overview,Shi2021DRFC,Tang2022MFRF}.
Since these functions are realized with a shared hardware and an integrated waveform, DFRC systems enjoy many advantages, such as reduced size and weight, improved hardware and spectral efficiency.

One important problem in DFRC systems is the proper design of transmit waveforms.
In \cite{Hassanien2016DFRC}, through controlling the sidelobes of the array beampattern by elaborate waveform design, the DFRC system was capable of detecting the target at the mainlobe and delivering the information bits toward the communication user at the sidelobes.
However, the bit rate associated with this approach was limited by the number of orthogonal waveforms.
In \cite{McCormick2017Simultaneous}, an algorithm was proposed to design waveforms for a MIMO array to simultaneously transmit radar and communication signals toward different directions.
In \cite{Liu2018DFRC,Tang2020DFRC,Shi2020DFRC}, the authors proposed waveform design method to match a radar beampattern and communicate with multiple users simultaneously.
In \cite{Liu2020Joint}, the authors proposed a joint transmit beamforming approach for DFRC systems.
The optimized waveforms therein could approximate a desired beampattern and guarantee the SINR performance for each communication user.
However, the designed waveforms in \cite{Liu2020Joint} are not constant-modulus (CM), which might result in nonlinear effects and distortions.

In this work, we investigate waveform design methods for DFRC systems in the presence of clutter.
To maximize the detection performance, the output signal-to-interference-plus-noise ratio (SINR) is used as the design metric.
Different from the waveform design approach in \cite{Tsinos2021DFRC}, which used an indirect method to control the overall synthesis error of the communication signals, we directly constrain the synthesis error associated with each communication user to be lower than a prescribed level (thus, the performance of every user is able to be controlled).
Moreover, we impose a CM constraint on the transmit waveforms, to comply the requirements of saturated amplifiers.
To tackle the encountered non-convex waveform design problem, a nested optimization algorithm is derived, which is based on cyclic optimization, Dinkinbach's transform, and alternating direction method of multipliers (ADMM).
Simulations show that the waveforms synthesized by the proposed algorithm can obtain superior detection performance and achieve multi-user communication capability at the same time.

The outline of the rest paper is as follows.
Section \ref{Sec:ProblemFormulation} presents the signal model and formulates the waveform design problem.
Section \ref{Sec:JointDesign} derives an iterative algorithm to tackle the waveform design problem.
Section \ref{Sec:NumericalResults} presents numerical examples to demonstrate the performance of the proposed algorithm and analyzes the impact of several factors (including the transmit energy of the desired communication signals, the number of communication users, the code length, and the synthesis errors) on the performance of DFRC systems.
Finally, conclusions are drawn in Section \ref{Sec:Conclusion}.

\emph{Notations}: The notations used in this paper are listed  in Table \ref{tab:notations}.
\begin{table}[!htbp]
\caption{{{List of Notations}}}
\renewcommand{\arraystretch}{1.15}
\centering
\begin{tabular}{cl}
	\toprule
	Symbol & Meaning\\
	\midrule
	$\bA$ & Matrix \\
	$\ba$ & Vector\\
	$a$ & Scalar\\
	$\bI$ & The identity matrix with the size determined by the subscript\\
	$(\cdot)^\ast$, $(\cdot)^\top$, $(\cdot)^\dagger$   & Conjugate, transpose, and conjugate transpose\\
	$(\cdot)^{1/2}$   & Squared root of a matrix\\
	$\textrm{tr}(\cdot)$ & Trace of a matrix\\
	$\left| \cdot \right|$, $\left\| \cdot \right\|_2$, $\left\| \cdot \right\|_\textrm{F}$ & Magnitude, Euclidian norm (of a vector), and Frobenius norm (of a matrix) \\
	$\expE\{\cdot \}$ & Expectation of a random variable\\
	$\realR$, $\complexC$ & Domain of real and complex numbers\\
	$\textrm{vec}(\cdot)$ & Vectorization\\
	$\otimes$ & Kronecker product\\
	$\textrm{Re}(\cdot)$ & The real part of a complex-valued scalar/vector/matrix\\
	$\bA \succ 0$ $(\bA \succeq 0)$ & $\bA$ is positive definite (semidefinite)\\
	\bottomrule
\end{tabular}
\label{tab:notations}
\end{table}

\section{Signal Model and Problem Formulation}\label{Sec:ProblemFormulation}
\begin{figure*}[!htbp]
\centering
\includegraphics[width= 0.55 \textwidth] {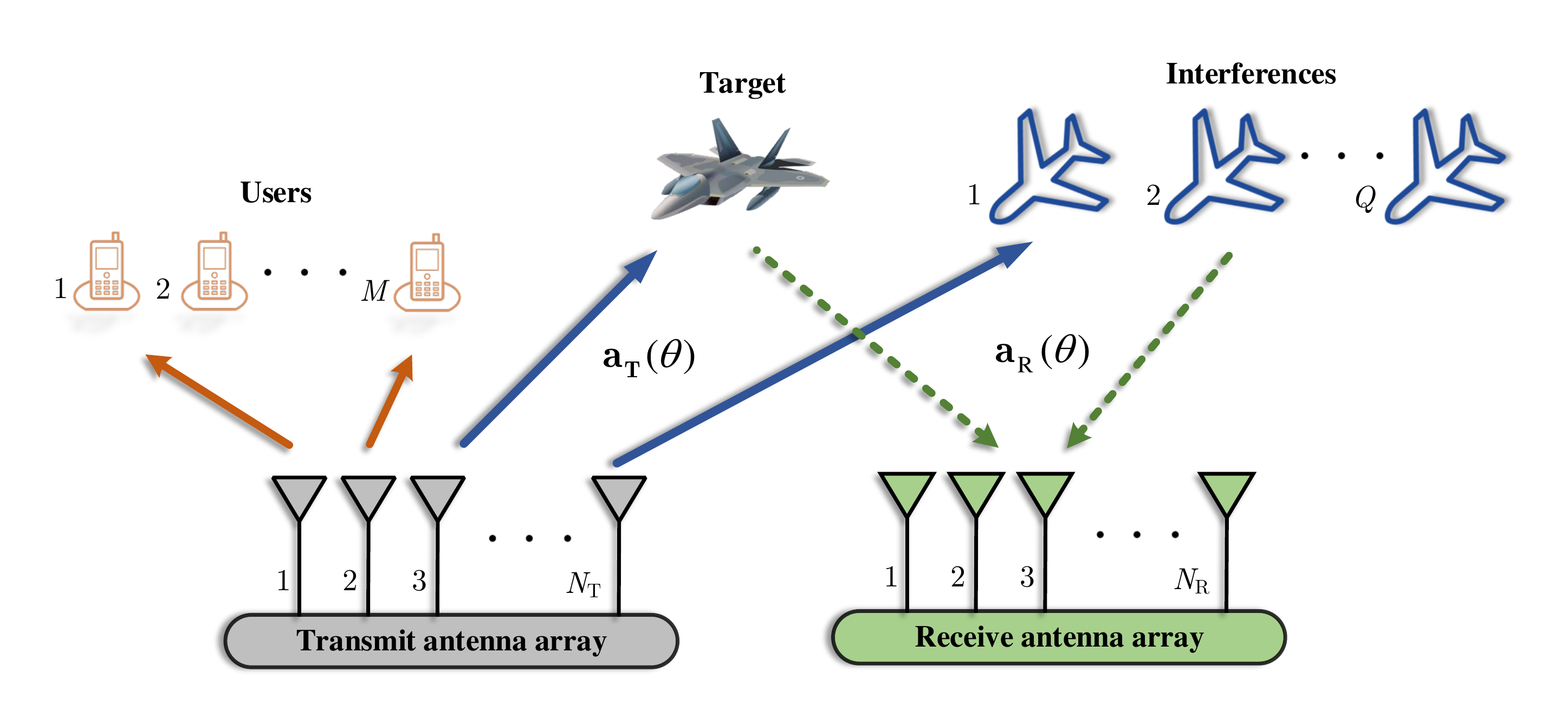}
\caption{Illustration of a MIMO DFRC system.}
\label{fig_1}
\end{figure*}

Consider a DFRC system based on a MIMO array, which has $N_{\textrm{T}}$ transmit and $N_{\textrm{R}}$ receive antennas, as illustrated in \figurename~\ref{fig_1}. We assume that a target of interest and $Q$ interference sources are present. In addition, the system serves $M$ communication users. To support simultaneous target detection and communication with multiple users, next we establish the signal models.

\subsection{Radar Model}	

Denote the target direction by $\theta_0$, and the direction of the $q$th interference source by $\theta_q$ ($\theta_q\neq\theta_0$, $q=1,2, \cdots, Q$).
The received signals at the $l$th instant ($l=1,2,\cdots,L$, and $L$ is the code length) is formulated by
\begin{align}\label{eq:sig_l}
\by_l &= \alpha_0 \ba_{\textrm{R}} (\theta_0) \ba_{\textrm{T}}^\top (\theta_0) \bx_l
+\sum\limits_{q=1}^Q \alpha_q \ba_{\textrm{R}} (\theta_q) \ba _{\textrm{T}}^\top(\theta_q) \bx_l + \bn_l,
\end{align}
where $\alpha_0, \alpha_1, \cdots, \alpha_Q$ are the amplitudes of the target and the $Q$ interference sources, $\ba_{\textrm{T}} (\theta)$ and $\ba_{\textrm{R}} (\theta)$ denote the transmit array steering vector and the receive array steering vector at $\theta$,  $\bx_l = [x_l(1), x_l(2), \cdots,$ $x_l({N_{\textrm{T}}})]^\top$,  $x_l(n)$ denotes the code of the $l$th subpulse of the (baseband) waveforms in the $n$th transmitter $(n = 1,2, \cdots N_{\textrm{T}})$, and $\bn_l$ is the receiver noise (we assume that it is white Gaussian).

Stacking $\by_1, \by_2,\cdots, \by_L$ in one column, we obtain
\begin{align} \label{eq:RadarSig}
\by = \alpha_0 \bA (\theta_0) \bx + \sum\limits_{q = 1}^Q \alpha_q \bA (\theta_q) \bx + \bn,
\end{align}
where $\by = [\by_1^\top,\by_2^\top,\cdots,\by_L^\top]^\top\in \complexC^{LN_{\textrm{R}} \times1}$,  $\bA (\theta) = \bI_L \otimes (\ba_{\textrm{R}} (\theta) \ba_{\textrm{T}}^\top (\theta))$, $\bx = [\bx_1^\top, \bx_2^\top,\cdots,\bx_L^\top]^\top\in \complexC^{LN_{\textrm{T}} \times1}$, and $\bn = [\bn_1^\top,\bn_2^\top,\cdots,\bn_L^\top]^\top\in \complexC^{LN_{\textrm{R}} \times1}$.

To detect the target, we pass $\by$ through a finite impulse response filter, denoted by $\bw$. The filter output $z$ can be written as
\begin{align}\label{eq:FilterOutput}
z &= \bw^\dagger \by \nonumber\\
&= \underbrace{\alpha_0 \bw^\dagger \bA (\theta_0) \bx}_{\textrm{Target}} + \underbrace{\bw^\dagger \sum\limits_{q = 1}^Q \alpha_q \bA (\theta_q) \bx}_{\textrm{Interference}} + \underbrace{\bw^\dagger \bn}_{\textrm{Noise}}.
\end{align}
According to \eqref{eq:FilterOutput}, we define the output SINR as follows:
\begin{align}\label{eq:SINR}
\textrm{SINR}_{\rm{r}} (\bx,\bw) =
\frac{\sigma_0^2|\bw^\dagger \bA (\theta_0) \bx|^2}{\bw^\dagger \left[ \sum\nolimits_{q = 1}^Q \sigma_q^2 \bA (\theta_q) \bx \bx^\dagger \bA^\dagger (\theta_q) \right]\bw + \sigma_{\textrm{n}}^2 \bw^\dagger \bw},
\end{align}
where the subscript r stands for ``radar", $\sigma_0^2$ is the target power, $\sigma_q^2 = \expE\{|\alpha_q|^2\}$ is the average power of the $q$th interference source, $q=1,2, \cdots, Q$, $\sigma _{\textrm{n}}^2$ is the noise power level, and we have assumed that the interference sources are uncorrelated.

\subsection{Communication Model}
The signals received by the $M$ users is given by
\begin{equation}\label{eq:CommSig}
\begin{split}
	\bY = \mathbf{HX} + \bZ,
\end{split}
\end{equation}
where $\bH=[\bh_1, \bh_2, \cdots, \bh_{M}]^\top \in \complexC^{M \times N_{\textrm{T}}}$ is the channel matrix; $\bX = [\bx_1, \bx_2, \cdots, \bx_L]\in \complexC^{  N_{\textrm{T}}\times L}$ denotes the transmit waveform matrix ($\bx = \textrm{vec}(\bX)$), and $\bZ$ is the noise matrix in the $M$ communication receivers. Let $\bs_m\in \complexC^{L \times 1}$ denote the desired symbols for the $m$th user ($m=1,2,\cdots,M$), and let $\bS = [\bs_1, \bs_2, \cdots, \bs_M]^\top\in\complexC^{M \times L}$.  Then we can rewrite \eqref{eq:CommSig} as
\begin{equation} \label{eq:CommSig2}
\begin{split}
	\bY = \bS+\underbrace{\mathbf{HX}-\bS}_{{\textrm{MUI}}}+ \bZ,
\end{split}
\end{equation}
where $\mathbf{HX}-\bS$ stands for the multi-user interference (MUI).
In \cite{Larsson2013precoding}, it is shown that for a Gaussian broadcasting channel and a Gaussian input, the achievable rate of the $m$th communication user ($m=1,2,\cdots,M$) is given by
\begin{equation}\label{eq:SumRate}
\begin{split}
	\ell_m = \log_2 (1 + {\textrm{SINR}}_{m,\rm{c}}),
\end{split}
\end{equation}
where ${\textrm{SINR}}_{m,\rm{c}}$ is the SINR of the $m$th communication receiver, defined by
\begin{equation}\label{eq:SINRComm}
\begin{split}
	{\textrm{SINR}}_{m,\rm{c}}= \frac{\expE\{ |s_{m,l}|^2\} }{\expE\{ |\bh_m^\top \bx_l - s_{m,l}|^2\} + \sigma_{z,m}^2},
\end{split}
\end{equation}
where the subscript c stands for ``communication", $s_{m,l}$ is the $l$th symbol of $\bs_m$ ($l=1,2,\cdots,L$), and $\sigma_{z,m}^2$ is the noise power level of the $m$th communication receiver.
Note that $\bh_m^\top \bx_l - s_{m,l}$ is the $(m,l)$th element of $\mathbf{HX}-\bS$. Thus, the minimization of
\begin{equation}\label{eq:MUI}
	\psi_m(\bX) = \| \bh_m^\top\bX - \bs_m^\top\|_2^2 =\| \bH_m\bx - \bs_m\|_2^2\approx L\expE\{ |\bh_m^\top \bx_l - s_{m,l}|^2\},
\end{equation}
which is the MUI for the $m$th communication receiver, results in the maximization of the achievable rate of the $m$th user, where $\bH_m=\bh_m^\top\otimes\bI_L$.

\subsection{Problem Formulation}
To maximize the target detection performance and ensure the communication performance of each user, we formulate the waveform design problem as follows:
\begin{align}
\max_{\bx,\bw}  &\  \textrm{SINR}_{\rm{r}} (\bx,\bw) \nonumber \\
\textrm{s.t.} &\ \psi_m(\bX) \leq \varsigma_m, m=1,2,\cdots,M, \nonumber \\
                 &\ \bx \in \mathcal{X},
\end{align}
where $\varsigma_m$ is the maximum allowed synthesis error of the $m$th user, and $\mathcal{X}$ denote the constraint set on the waveforms. In practice, to make the radio frequency (RF) amplifier work at maximum efficiency and avoid nonlinear distortions, CM waveforms are often used. Thus, we enforce the CM constraint on the transmit waveforms, i.e.,
\begin{equation}\label{eq:CM}
	\begin{split}
		|\bx(n)| = \sqrt {p_s} , n=1, 2, \cdots,  LN_{\textrm{T}},
	\end{split}
\end{equation}
where $\bx(n)$ indicates the $n$th element of $\bx$, $p_s = e_{\textrm{T}} /(LN_{\textrm{T}})$, and $e_{\textrm{T}}$ is the total transmit energy.

By combining the results in \eqref{eq:SINR}, \eqref{eq:MUI}, and \eqref{eq:CM}, the joint design of transmit waveforms and receive filters for the DFRC system can be formulated as
\begin{align} \label{eq:objective function}
\max_{\bx,\bw}  &\  \frac{|\bw^\dagger \bA (\theta_0) \bx|^2}{\bw^\dagger \left[ \sum\nolimits_{q = 1}^Q \sigma_q^2 \bA (\theta_q) \bx \bx^\dagger \bA^\dagger (\theta_q) \right]\bw + \sigma_{\textrm{n}}^2 \bw^\dagger \bw} \nonumber \\
\textrm{s.t.} &\ \|  \bH_m\bx - \bs_m\|_2^2 \leq \varsigma_m, m=1, 2,\cdots,M, \nonumber \\
                 &\ |\bx(n)| = \sqrt {p_s} , n=1, 2, \cdots,  LN_{\textrm{T}}.
\end{align}

\textit{Remark 1:} In the formulation of \eqref{eq:objective function}, we have assumed that the average power and the directions of the interference sources as well as the channel matrix are known \textit{a priori}. Such an assumption is justified if the access to the estimates of them from the previous scans are available (see also in \cite{Tang2020Polyphase,Tsinos2021DFRC} for similar assumptions).

\textit{Remark 2:} We note that the formulated problem in \eqref{eq:objective function} is different from that in \cite{Tsinos2021DFRC}. The studies in \cite{Tsinos2021DFRC} controlled the total MUI energy indirectly by tuning the combination coefficient, while we constrain the MUI energy of each user directly to control their communication performance. More specifically, according to \eqref{eq:MUI}, if the per-user MUI energy constraint is satisfied, we can approximately obtain that
\begin{align}
	\expE\{ |\bh_m^\top \bx_l - s_{m,l}|^2\} \leq \varsigma_m/L, m=1,2,\cdots,M.
\end{align}
As a result, the achievable information rate for the $m$th user satisfies
\begin{align}
	\ell_m \geq  \log_2 \left(1 + \frac{p_m}{\varsigma_m/L+\sigma_{z,m}^2}\right),
\end{align}
where $p_m = \expE\{ |s_{m,l}|^2\}$ is the average power of the $m$th communication signals, $m=1,2,\cdots,M$.
On the other hand, if only the total MUI energy constraint is enforced, the initialization of the transmit waveforms will affect the communication performance of each user (i.e., the communication performance of each user cannot be guaranteed).

\section{Algorithm Design}\label{Sec:JointDesign}
Due to the CM constraint, it is evident that the fractional programming problem in \eqref{eq:objective function} is non-convex. In this section, a cyclic optimization method is used to deal with this non-convex problem. Specifically, at the $(k+1)$th iteration, we optimize $\bw^{(k+1)}$ for fixed $\bx^{(k)}$; then we optimize $\bx^{(k+1)}$ for fixed $\bw^{(k+1)}$. Next we present solutions to the two optimization problems. To lighten the notations, we omit the superscript if doing so does not result in confusions.
\subsection{Optimization of $\bw^{(k+1)}$ for fixed $\bx^{(k)}$}
The corresponding optimization problem is given by
\begin{align} \label{eq:ProblemFixedS}
\max_{\bw}  &\  \frac{|\bw^\dagger \bA (\theta_0) \bx|^2}{\bw^\dagger \left[ \sum\nolimits_{q = 1}^Q \sigma_q^2 \bA (\theta_q) \bx \bx^\dagger \bA^\dagger (\theta_q) \right]\bw + \sigma_{\textrm{n}}^2 \bw^\dagger \bw}.
\end{align}
It can be checked that the solution to \eqref{eq:ProblemFixedS} is given by
\begin{equation}\label{eq:slove w}
	\begin{split}
		\bw= \gamma \bR_x^{-1} \bA (\theta_0) \bx,
	\end{split}
\end{equation}
where 	$\gamma \neq 0$ is an arbitrary constant, and
\begin{equation}\label{eq:transform 1}
	\begin{split}
		\bR_x = \sum\limits_{q = 1}^Q \sigma_q^2 \bA (\theta_q) \bx \bx^\dagger \bA^\dagger (\theta_q) + \sigma_{\textrm n}^2 \bI_{LN_{\textrm{R}}}.
	\end{split}
\end{equation}

\subsection{Optimization of $\bx^{(k+1)}$ for fixed $\bw^{(k+1)}$}
Define
\begin{equation}\label{eq:R0}
		\bR_0= \bA^\dagger (\theta_0) \bw\bw^\dagger \bA (\theta_0)
\end{equation}
and
\begin{equation} \label{eq:R1}
		\bR_1 = \sum\limits_{q = 1}^Q \sigma_q^2 \bA^\dagger (\theta_q) \bw \bw^\dagger \bA (\theta_q)  + \sigma_{\textrm n}^2 \bw^\dagger \bw /e_{\textrm{T}}\cdot \bI_{LN_{\textrm{T}}}.
\end{equation}
Then the waveform design problem for fixed $\bw^{(k+1)}$ can be written as
\begin{align} \label{eq:ProblemFixedW}
\max_{\bx}  &\   \frac{\bx^\dagger \bR_0 \bx }{\bx^\dagger \bR_1 \bx} \nonumber \\
\textrm{s.t.} &\ \| \bH_m\bx - \bs_m\|_2^2 \leq \varsigma_m, m=1,2,\cdots,M, \nonumber \\
                 &\ |\bx(n)| = \sqrt {p_s} , n=1, 2, \cdots,  LN_{\textrm{T}}.
\end{align}

Next we resort to Dinkelbach's transform \cite{1967Dinkelbach} to deal with the optimization problem in \eqref{eq:ProblemFixedW}. To this end, we use $g(\bx^{(k,l)})$ to indicate the objective value of \eqref{eq:ProblemFixedW} at the $(k,l)$th iteration (Here we use the superscript $k$ to indicate the outer iteration for cyclic optimization, and $l$ to denote the inner iteration for  Dinkelbach's transform). By using Dinkelbach's transform, the optimization problem at the $(k,l+1)$th iteration is formulated as
\begin{align} \label{eq:DinT}
\max_{\bx}  &\    \bx^\dagger \widetilde{\bT} \bx \nonumber \\
\textrm{s.t.} &\ \|  \bH_m\bx - \bs_m\|_2^2 \leq \varsigma_m, m=1,2, \cdots,M, \nonumber \\
                 &\ |\bx(n)| = \sqrt {p_s} , n=1, 2, \cdots,  LN_{\textrm{T}},
\end{align}
where $\widetilde{\bT}= \bR_0 - g(\bx^{(k,l)})\bR_1$. To proceed, we define
\begin{equation} \label{eq:T}
\bT= \widetilde{\bT} - \beta \bI_{LN_{\textrm{T}}},
\end{equation}\label{eq:proof T}where $\beta \leq \lambda_{\textrm{min}}(\widetilde{\bT})$, and $\lambda_{\textrm{min}}(\widetilde{\bT})$ is the smallest eigenvalue of $\widetilde{\bT}$. It is evident that $\bT$ is positive semidefinite such that its square root exists. In addition,
\begin{equation}
\bx^\dagger {\bT}\bx = \bx^\dagger \widetilde{\bT}\bx - \beta e_{\textrm{T}}.
\end{equation}
Therefore, the optimization problem in \eqref{eq:DinT} is equivalent to
\begin{align} \label{eq:DinT2}
\max_{\bx}  &\    \bx^\dagger {\bT}  \bx \nonumber \\
\textrm{s.t.} &\ \|  \bH_m\bx - \bs_m\|_2^2 \leq \varsigma_m, m=1,2,\cdots,M, \nonumber \\
                 &\ |\bx(n)| = \sqrt {p_s} , n=1, 2, \cdots,  LN_{\textrm{T}}.
\end{align}
Next we apply ADMM (we refer to \cite{Boyd2010ADAMM} for a comprehensive survey of ADMM) to tackle \eqref{eq:DinT2}. By using the variable splitting trick and introducing auxiliary variables $\hbx$ and $\tbx_m$ ($m=1,\cdots,M$), the optimization problem in \eqref{eq:DinT2} is recast by
\begin{subequations}\label{eq:objective function ADMM}
\begin{align}
  \max_{\bx,\hbx,\tbx}   &\    \hbx^\dagger \hbx  \\
  \textrm{s.t.}  &\ \hbx = \bT^{1/2} \bx,  \label{eq:constraintA}\\
                     &\ \| \tbx_m\|_2^2 \leq \varsigma_m, \tbx_m = \bH_m\bx - \bs_m, m=1,2, \cdots,M,  \label{eq:constraintB}\\
                     &\ |\bx(n)| = \sqrt {p_s} , n=1, 2, \cdots,  LN_{\textrm{T}}. \label{eq:constraintC}
\end{align}
\end{subequations}

The augmented Lagrange function of \eqref{eq:objective function ADMM} is written as
\begin{align}\label{eq:Lagrange function}
	\mathop{{L_\mu }(\bx,\hbx,\{\tbx_m\},\bnu,\{\ups_m\})} &=  - \hbx^\dagger \hbx\nonumber\\			
	&+ \frac{\mu}{2}\left[ {\| \hbx -\bT^{1/2}\bx + \bnu \|_2^2 - \left\| \bnu \right\|_2^2} \right]\nonumber\\
	&+ \frac{\mu}{2}\left\{\sum\limits_{m = 1}^M \Big{[} {\| \tbx_m - \bH_m \bx + \bs_m + \ups_m \|_2^2 - \| \ups_m \|_2^2} \Big{]} \right\},
\end{align}
where ${\mu}$ is the penalty parameter; $\bnu$ and $\ups_m (m=1,\cdots,M)$ are the Lagrange multipliers associated with the constraints in \eqref{eq:constraintA} and \eqref{eq:constraintB}, respectively. In the $(t+1)$th iteration of the ADMM algorithm, we carry out the following steps sequentially:
\begin{equation}\label{eq:ADMM1}
\bx^{{(t+1)}} = \mathop {\rm{argmin}} \limits_{\bx \in \mathcal{X}} L_\mu (\bx,  \hbx^{(t)}, \{\tbx_m^{(t)}\},  \bnu^{(t)}, \{\ups_m^{(t)}\}),
\end{equation}	
\begin{equation}\label{eq:ADMM2}
\hbx^{{(t+1)}}  = \mathop {\rm{argmin}} \limits_{\hbx} L_\mu (\bx^{{(t+1)}}, \hbx, \{\tbx_m^{(t)}\}, \bnu^{(t)}, \{\ups_m^{(t)}\}),
\end{equation}	
\begin{equation}\label{eq:ADMM3}
\tbx_m^{{(t+1)}} = \mathop {\rm{argmin}} \limits_{\tbx_m} L_\mu (\bx^{{(t+1)}}, \hbx^{{(t+1)}}, \{\tbx_m\}, \bnu^{(t)}, \{\ups_m^{(t)}\}),
\end{equation}	
\begin{equation}\label{eq:ADMM4}
\bnu^{{(t+1)}} = \bnu^{(t)} + \hbx^{{(t+1)}} - \bT^{1/2} \bx^{{(t+1)}},
\end{equation}	
\begin{equation}\label{eq:ADMM5}
\ups_m^{{(t+1)}} = \ups_m^{(t)} + \tbx_m^{{(t+1)}} - \bH_m \bx^{{(t+1)}} + \bs_m.
\end{equation}

Next we derive the solutions to \eqref{eq:ADMM1}, \eqref{eq:ADMM2}, and \eqref{eq:ADMM3}.

\subsubsection{Solution to \eqref{eq:ADMM1}\rm}
The optimization problem in \eqref{eq:ADMM1} can be recast as
\begin{align}\label{eq:Update1}
\min_{\bx}&\  \| \hbx - \bT^{1/2}\bx + \bnu \|_2^2 + \sum\limits_{m = 1}^M\| \tbx_m - \bH_m \bx + \bs_m + \ups_m \|_2^2 \nonumber \\
\textrm{s.t.} &\ |\bx(n)| = \sqrt {p_s} , n=1, 2, \cdots,  LN_{\textrm{T}}.
\end{align}
Let
\begin{equation}\label{eq:B}
\bB = \bT + \sum\limits_{m = 1}^M\bH_m^\dagger \bH_m,
\end{equation}
and
\begin{equation}\label{eq:b}
\bb = \bT^{1/2}(\hbx+\bnu)+\sum\limits_{m = 1}^M\bH_m^\dagger(\tbx_m + \bs_m + \ups_m).
\end{equation}
Then, we can rewrite the optimization problem in \eqref{eq:Update1} as
\begin{align}\label{eq:ADMM1 transform}
\mathop {\min}\limits_\bx &\ \bx^\dagger \bB \bx - 2{\mathop{\textrm{Re}}\nolimits} (\bb^\dagger \bx) \nonumber \\
\textrm{s.t.} &\ |\bx(n)| = \sqrt {p_s} , n=1, 2, \cdots,  LN_{\textrm{T}}.
\end{align}
Note that \eqref{eq:ADMM1 transform} is a standard unimodular quadratic programming (UQP) problem. A number of algorithms have been proposed to tackle the UQP problem, including the power-method like (PML) iterations \cite{Soltanalian2013Joint,Soltanalian2014Optimization}, the coordinate descent method (CDM) \cite{Cui2017Quadratic,Tsinos2022CM}, the gradient projection (GP) method \cite{Tranter2017CM}, and the majorization-minimization (MM) method \cite{Tang2019Efficient,Tang2021Information,Tang2021Profiling,Tsinos2022CM,Palomar2018CM}. We find that these methods have similar performance but the MM method is usually the fastest. Therefore, we use the MM method to tackle the problem in \eqref{eq:ADMM1 transform}.

\subsubsection{Solution to \eqref{eq:ADMM2}\rm}
The optimization problem in \eqref{eq:ADMM2} is formulated by
\begin{equation}\label{eq:Update2}
\min_{\hbx}  - \hbx^\dagger \hbx + \frac{\mu}{2}\| \hbx -\bT^{1/2} \bx + \bnu \|_2^2 .
\end{equation}
Define
\begin{equation}\label{eq:q}
\bq = \bT^{1/2} \bx -\bnu.
\end{equation}
Then, we can recast \eqref{eq:Update2} as
\begin{equation}\label{eq:ADMM2 transform}
\mathop {\min}\limits_\hbx \frac{\mu}{2}\left\| \hbx - \bq \right\|_2^2 - \hbx^\dagger \hbx,
\end{equation}
Assume that $\mu >2$. Then the quadratic optimization problem in \eqref{eq:ADMM2 transform} is convex. Taking the derivative of \eqref{eq:ADMM2 transform} with respect to $\hbx$ and setting it equal to zero, we can acquire the solution to \eqref{eq:ADMM2 transform}:
\begin{equation}\label{eq:solve ADMM2 transform}
\hbx= \frac{\mu}{\mu-2} \bq.
\end{equation}

\subsubsection{Solution to \eqref{eq:ADMM3}\rm}
Define
\begin{equation}\label{eq:p}
\bp_m=\bH_m \bx- \ups_m - \bs_m.
\end{equation}
Then, we can rewrite \eqref{eq:ADMM3} as
\begin{align} \label{eq:ADMM3 transform}
\min_{\tbx_m}  &\  \left\| \tbx_m- \bp_m \right\|_2^2 \nonumber \\
\textrm{s.t.} &\ \| \tbx_m\|_2^2 \leq \varsigma_m.
\end{align}
It can be verified that the solution to \eqref{eq:ADMM3 transform} is
\begin{align}\label{eq:slove ADMM3 transform}
\tbx_m =
\begin{cases}
\bp_m, & \|\bp_m\|_2^2 \le \varsigma_m, \\
\sqrt{\varsigma_m} \bp_m/\left\| \bp_m \right\|, & \|\bp_m\|_2^2 > \varsigma_m.
\end{cases}
\end{align}

We summarize the proposed ADMM algorithm in Algorithm \ref{alg1}, where we stop the proposed ADMM method when the norm of the primal residual $\|\br_m^{(t)}\|_2 \leq \epsilon^{\textrm{primal}}$ ($m = 1,2,\cdots,M+1$) and the norm of the dual residual
$\|\bd_m^{(t)}\|_2 \leq \epsilon^{\textrm{dual}}$ ($m = 1,2,\cdots,M+2$), where
\begin{align}\label{eq:primal residual}
\br_m^{(t)}=
\begin{cases}
\bH_m \bx^{(t)} - \bs_m - \tbx_m^{(t)}, &1\leq m \leq M,\\
\bT^{1/2} \bx^{(t)} - \hbx^{(t)}, &m = M+1,
\end{cases}
\end{align}
\begin{align}\label{eq:Dual residuals}
\bd_m^{(t)} =
\begin{cases}
  \tbx_m^{(t)} -  \tbx_m^{(t-1)}, &  1\leq m \leq M,\\
  \hbx^{(t)} - \hbx^{(t-1)}, &m = M+1, \\
  \bx^{(t)} - \bx^{(t-1)}, &m = M+2,
\end{cases}
\end{align}
${\epsilon^{\textrm{primal}}} > 0$ and ${\epsilon^{\textrm{dual}}} > 0$ are the feasible tolerances of the primal and dual conditions, respectively.
\begin{algorithm}[!htbp]
\renewcommand{\arraystretch}{1.3}
  \caption{ \small  ADMM algorithm for the problem in \eqref{eq:DinT2}.}\label{alg1}
  \KwIn{$\bT, p_s,\{\bH_m, \bs_m,\varsigma_m\}_{m=1}^M$.}
  \KwOut{$\bx^{(k,l+1)}$.}
  \textbf{Initialize:} \\
  Compute $\bB$ by \eqref{eq:B}. \\
  $t=0,\bx^{(t)}=\bx^{(k,l)}$, $\bnu^{(t)} = \bzero$, $\ups_m^{(t)} = \bzero$.\\
  Compute $\hbx^{(t)}$, $\{\tbx_m^{(t)}\}_{m=1}^M$. \\
    \Repeat{convergence}{
    $\bb^{(t)} = \bT^{1/2}(\hbx^{(t)}+\bnu^{(t)})+\sum\nolimits_{m = 1}^M\bH_m^\dagger(\tbx_m^{(t)} + \bs_m + \ups_m^{(t)}).$\\
    Update $\bx^{{(t+1)}}$ through solving  \eqref{eq:ADMM1 transform} by MM.\\
    $\bq^{(t)} = \bT^{1/2} \bx^{{(t+1)}} -\bnu^{{(t)}}$.\\
    $\hbx^{(t+1)} = \dfrac{\mu}{\mu-2} \bq^{(t)} $. \\
    $\bp_m^{(t)} =\bH_m \bx^{(t+1)} - \ups_m^{(t)}  - \bs_m$. \\
    $\tbx_m^{(t+1)} = \min(\sqrt{\varsigma_m} /\| \bp_m^{(t)}  \|,1)\cdot \bp_m^{(t)} $.\\
    $\bnu^{{(t+1)}} = \bnu^{(t)} + \hbx^{{(t+1)}} - \bT^{1/2} \bx^{{(t+1)}}$. \\
    $\ups_m^{{(t+1)}} = \ups_m^{(t)} + \tbx_m^{{(t+1)}} - \bH_m \bx^{{(t+1)}} + \bs_m.$ \\
    $t = t+1$.\\
    }
   $\bx^{(k,l+1)}=\bx^{(t)}$.
\end{algorithm}

\subsection{Algorithm Summary and Computational Complexity Analysis}
We summarize the proposed constant-modulus waveform design algorithm for the DFRC systems in Algorithm \ref{alg2}, where we terminate the algorithm if
\begin{equation}\label{eq:Stop Criterion}
\frac{|{\textrm{SINR}^{(k)}_{\rm{r}}}-{\textrm{SINR}^{(k-1)}_{\rm{r}}}|}{\textrm{SINR}^{(k-1)}_{\rm{r}}} < \vartheta_{\textrm{O}},
\end{equation}
and we terminate the inner loop (for Dinkelbach's transform) if
\begin{equation}\label{eq:inner loop Criterion}
\frac{|g(\bx^{(k,l)})-g(\bx^{(k,l-1)})|}{g(\bx^{(k,l-1)})}< \vartheta_{\textrm{I}},
\end{equation}
where $\vartheta_{\textrm{O}}$  and $\vartheta_{\textrm{I}}$ are predefined small values (e.g., $10^{-5}$).

\begin{algorithm}[!htbp] \label{alg2}
\caption{\small Joint Design of CM Waveforms and Receive Filters for DFRC Systems}
  \KwIn{$\left\{\theta_q, \sigma_q^2 \right\}_{q = 1}^Q, \theta_0, \alpha_0^2,  \{\bs_m,\varsigma_m\}_{m=1}^M, \bH, p_s$}
  \KwOut{$\bx$, $\bw$}
  \textbf{Initialize:} \\
  $\left\{\bH_m =  \bh_m^{\top} \otimes \bI_L\right\}_{m = 1}^M$, $\bA(\theta) = \bI_L \otimes (\ba_{\textrm{R}}(\theta)\ba_{\textrm{T}}^\top(\theta))$.\\
  $k=0, \bx^{(0)}$.\\
    \Repeat{convergence}{   	
    	$\textit {// Optimization of}$ $\bw$.\\
    	Compute $\bR_x^{(k)}$ by \eqref{eq:transform 1}.\\
    	Update $\bw^{(k+1)}$ by \eqref{eq:slove w}.\\
    	$\textit {// Optimization of}$ $\bx$.\\
    	Compute $\bR_0^{(k)}$ by \eqref{eq:R0}.\\
    	Compute $\bR_1^{(k)}$ by \eqref{eq:R1}.\\
    	$l=0, \bx^{(k,l)}=\bx^{(k)}$.\\
    	\Repeat{convergence}{
    		Compute $g(\bx^{(k,l)})=  \dfrac{(\bx^{(k,l)})^\dagger \bR_0^{(k)} \bx^{(k,l)} }{(\bx^{(k,l)})^\dagger \bR_1^{(k)} \bx^{(k,l)}}$.\\
    		Compute $\tbT^{(k,l)} = \bR_0^{(k)}-g(\bx^{(k,l)})\bR_1^{(k)}$.\\
    		Compute $\bT^{(k,l)}= \widetilde{\bT}^{(k,l)} - \beta \bI_{LN_{\textrm{T}}}$.\\
    		Compute $\bx^{(k,l+1)}$ by Algorithm \ref{alg1}.\\
    		$l=l+1$.\\
    	}
     $\bx^{(k+1)}=\bx^{(k,l)}$.\\
     $k=k+1$.\\   	
    }
    $\bx=\bx^{(k)},\bw=\bw^{(k)}$.
\end{algorithm}

The computational complexity of Algorithm \ref{alg2} is determined by the number of outer iterations and the complexity at each outer iteration. We present the computational complexity for each outer iteration in Table \ref{tab:Complexity}, where $N_{\textrm{D}}$ and $N_{\textrm{A}}$ denote the number of iterations for Dinkelbach's transform and the proposed ADMM algorithm to reach convergence, respectively.

\textit{Remark 3:} Due to the constant-modulus constraint and the fractional objective function, the optimization problem in \eqref{eq:objective function} is non-convex and difficult to tackle. To synthesize the constant-modulus waveforms, we derive Algorithm \ref{alg2}, which is a nested optimization algorithm based on the cyclic optimization, Dinkinbach's transform, and ADMM. Essentially, cyclic optimization and the method based on Dinkinbach's transform are ascent algorithms. However, the convergence property of the proposed ADMM algorithm, which deals with a non-convex optimization problem, remains unknown (for some recent progress on this problem, we refer to \cite{Hong2016Convergence}). Therefore, it is non-trivial to prove the convergence property of Algorithm \ref{alg2}. Fortunately, we have not encountered any convergence problems in our extensive numerical studies.

\begin{table}[!htbp]
	\caption{{{Computational Complexity Analysis}}}
	\renewcommand\arraystretch{1.2}
	\centering
	\begin{tabular}{l|ll}
		\toprule
		\multicolumn{2}{c}{Computation}  & {Complexity} \\
		\cline{1-3}
		\multirow{3}{*}{Update $\bw$} & $\bR_x$ & $O(L^2 N_{\textrm{R}} N_{\textrm{T}})$ \\
		\multirow{3}{*}{} & $\bR_x^{-1}$ & $O(L^3 N_{\textrm{R}}^3)$\\
		\multirow{3}{*}{} & $\bw$ & $O(L^2 N_{\textrm{R}} N_{\textrm{T}})$\\
		\cline{1-3}
		\multirow{3}{*}{Update $\bx$} & $\bR_0, \bR_1$ & $O(L^2 N_{\textrm{R}} N_{\textrm{T}})$ \\
		\multirow{3}{*} & $g(\bx)$ & $O(N_{\textrm{D}}L^2 N_{\textrm{T}}^2)$ \\
		\multirow{3}{*} & Compute $\bx^{(k,l+1)}$ by Algorithm 1 & $O(N_{\textrm{D}}N_{\textrm{A}}(2M+3)L^2 N_{\textrm{T}}^2)$ \\
		\cline{1-3}
		Total & \multicolumn{2}{c}{$O(L^3 N_{\textrm{R}}^3+O(N_{\textrm{D}}N_{\textrm{A}}(2M+3)L^2 N_{\textrm{T}}^2)$} \\
		\cline{1-3}
		
	\end{tabular}
	\label{tab:Complexity}
\end{table}

\renewcommand\arraystretch{1}

\subsection{Algorithm Extension}
We note that the proposed algorithm can be extended to deal with the peak-to-average-power-ratio (PAPR) constraint. To this purpose, we only need to replace the optimization problem in \eqref{eq:ADMM1 transform} by the following:
\begin{align}\label{eq:ADMM_PAPR}
	\mathop {\min}\limits_\bx &\ \bx^\dagger \bB \bx - 2{\mathop{\textrm{Re}}\nolimits} (\bb^\dagger \bx) \nonumber \\
	\textrm{s.t.} &\ \bx(n)^\dagger\bx(n) = {e_{\textrm{T}}}/N_\textrm{T}, \textrm{PAPR}(\bx(n)) \le \rho, n=1,2,\cdots,N_\textrm{T}.
\end{align}
where $\bx(n)$ denotes the waveforms of the $n$th transmitter, $1 \le \rho \le L$,
\begin{align}\label{eq:PAPR}
	\textrm{PAPR}(\bx(n)) = \dfrac{\max_l |\bx_l(n)|^2}{{\dfrac{1}{L}}\sum\nolimits_{l = 1}^{L} |\bx_l(n)|^2},
\end{align}
and we have assumed that the transmit energy across the antenna elements are uniform. Similarly, we can use the MM method to tackle the optimization problem in \eqref{eq:ADMM_PAPR}  \cite{Tang2021Information}. In addition to the PAPR constraint, we can also extend the proposed algorithm to design waveforms under a similarity constraint \cite{Li2006SPL,Maio2008TSP} as well as both similarity and constant-modulus constraints. However, we skip the details due to space limitations.

\section{Numerical Results}\label{Sec:NumericalResults}
In this section, numerical examples are provided to demonstrate the performance of the proposed algorithm.
The DFRC system under consideration has $N_{\textrm{T}}=16$  transmit antennas and $N_{\textrm{R}}=8$ receive antennas.
Both the transmit and the receive antenna array are uniform linear arrays, with inter-element spacing being $\lambda/2$ ($\lambda$ is the wavelength).
The target is at the direction of $\theta_0 = 20^\circ$ and has a power of $0$ dB.
We assume that $Q=4$ interference sources are present, with the directions and power being $\left\{-40^\circ, -20^\circ, 40^\circ, 50^\circ \right\}$ and $\sigma_q^2 = 30$ dB ($q = 1,2,3,4$), respectively.
The noise power level in the radar receiver is $0$ dB.
The available transmit energy is ${e_{\textrm{T}}} = 20$.
The elements of the channel matrix $\bH$ are independent and identically distributed, obeying a Gaussian distribution with zero mean and variance of $1$ (i.e., we assume a flat fading channel).
Regarding the stopping criteria of the proposed algorithm, we set ${\epsilon^{\textrm{primal}}}=10^{-4}$, ${\epsilon^{\textrm{dual}}}=10^{-2}$, and $\vartheta_{\textrm{I}}=\vartheta_{\textrm{O}}= 10^{-5}$.
Finally, all the analysis is performed on a standard laptop with CPU CoRe i7 2.8 GHz and 16 GB of RAM.

\begin{figure*}[!htbp]
	\centering
	\includegraphics[width=2.9in]{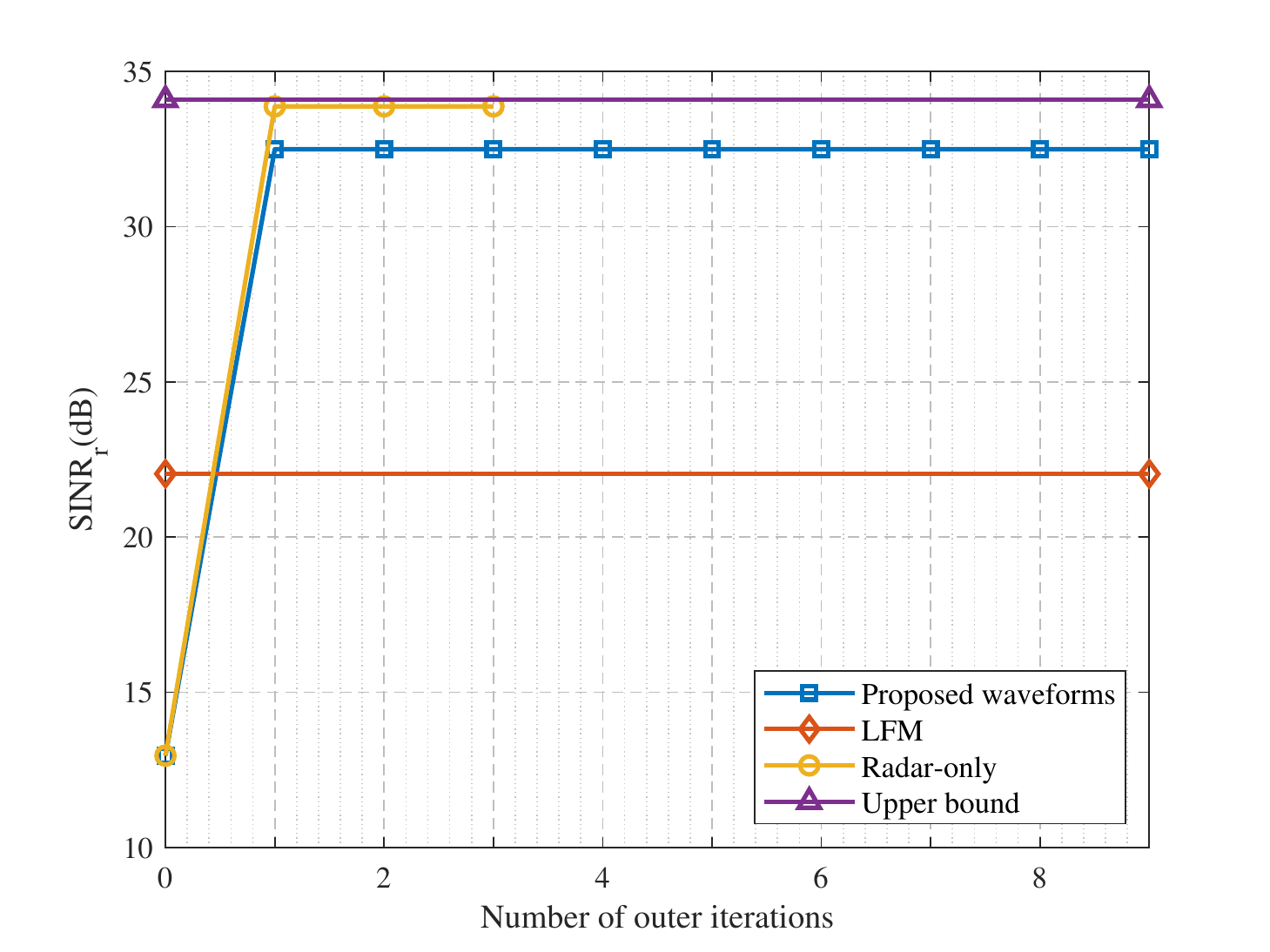}
	\caption{$\textrm{SINR}_{\rm{r}}$ versus the number of outer iterations. $M = 2, e_1 = e_2 = 20, L =20, \varsigma_1 = 10^{-3}, \varsigma_2 = 5\times 10^{-3}$.}
	\label{fig_2}
\end{figure*}

Firstly, we analyze the convergence of the proposed algorithm. We assume $M=2$ communication users. The desired signals of user 1 and user 2 have a quadrature phase shift keying (QPSK) and an 8-quadrature amplitude modulation (8QAM) modulation, respectively. The associated information bits are randomly generated. The transmit energies of the desired communication signals are $e_1=e_2=20$. \figurename~\ref{fig_2} illustrates the convergence of the $\textrm{SINR}$ of the synthesized waveforms versus the number of outer iterations, where the code length is $L = 20$, and the maximum allowed synthesis errors are $ \varsigma_1 = 10^{-3}$ and  $\varsigma_2 = 5\times 10^{-3}$, respectively. In addition, the upper bound\footnote{It can be verified that the upper bound of $\textrm{SINR}_{\rm{r}}$ is $N_{\textrm{T}}N_{\textrm{R}}{e_{\textrm{T}}}$.} and the $\textrm{SINR}$ curve associated with the radar-only case\footnote{The waveforms for the radar-only case are synthesized by removing the communication constraints in \eqref{eq:objective function}, and the associated waveform design problem can be tackled by Algorithm 3 in \cite{Tang2016TxRx}.} are also drawn. We can see that the $\textrm{SINR}$ at convergence is $32.49$ dB, which is about $10$ dB higher than that of the LFM signals. However, since the DFRC system has to spare some transmit energy to satisfy the additional communication constraints, the proposed waveforms will suffer some performance loss (compared with the upper bound and the radar-only case, the loss of the proposed waveforms are about $1.59$ dB and $1.38$ dB, respectively).

\begin{figure*}[!htbp]
	\centering
	\includegraphics[width=2.9in]{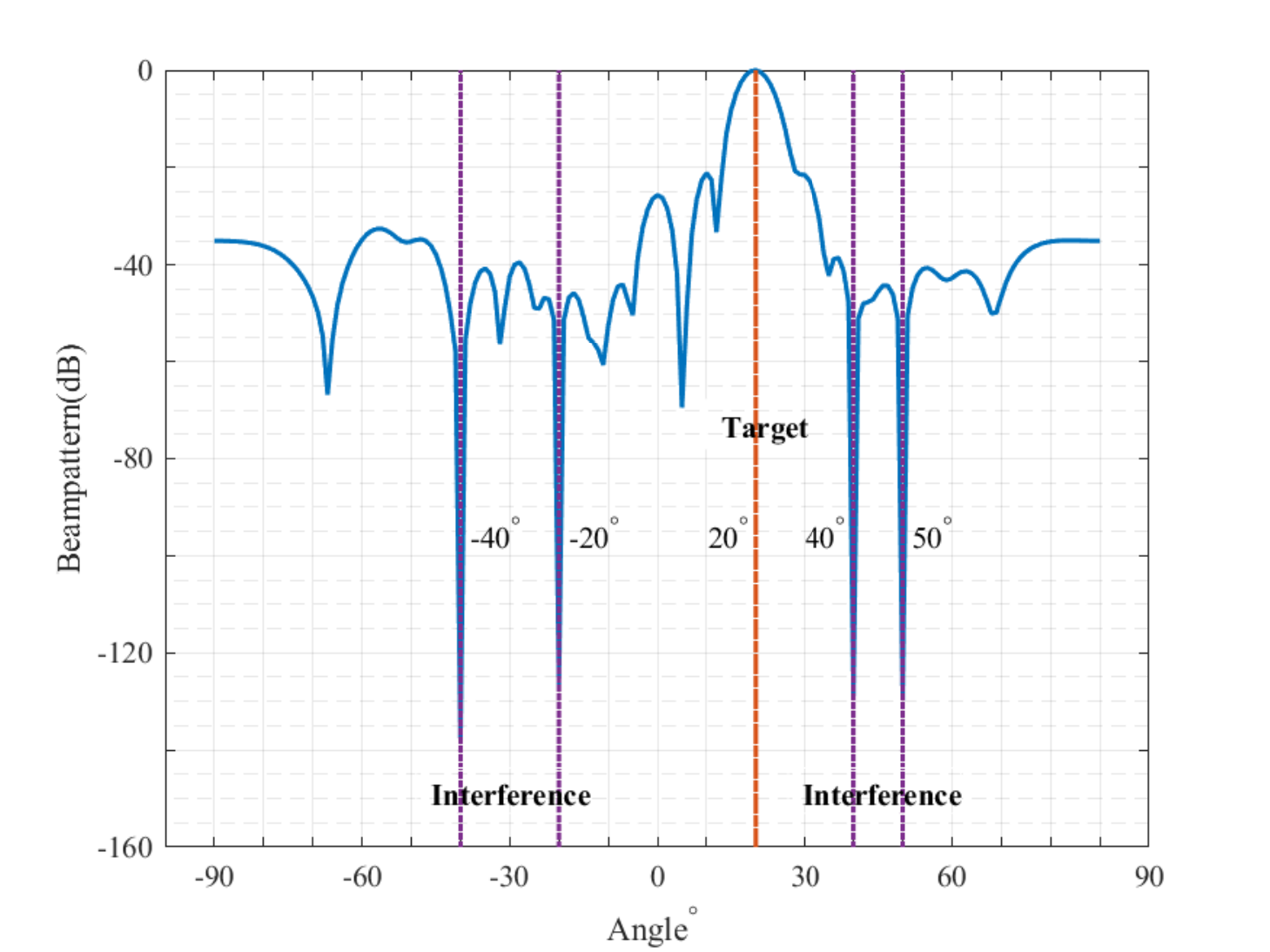}
	\caption{Beampattern of the DFRC systems. $M = 2, e_1 = e_2 = 20, L =20, \varsigma_1 = 10^{-3}, \varsigma_2 = 5\times 10^{-3}$.}
	\label{fig_3}
\end{figure*}

\figurename~\ref{fig_3} shows the beampattern associated with the designed waveforms and filters, where the beampattern is defined by
\begin{equation}\label{eq:pattern}
	\begin{split}
		P(\theta) = |\bw^\dagger \bA(\theta) \bx|^2.
	\end{split}
\end{equation}
Note that the beampattern has the highest gain at the target direction ($\theta_0 = 20^\circ$), and forms deep nulls (lower than $-120$ dB) at the interference directions ($-40^\circ, -20^\circ, 40^\circ,$ and $50^\circ $). Therefore, the designed transmit waveforms and receive filters can suppress the interference power to a very low level, ensuring the target detection performance.

\begin{figure*}[!htbp]
	\centering
	\includegraphics[width=2.9in]{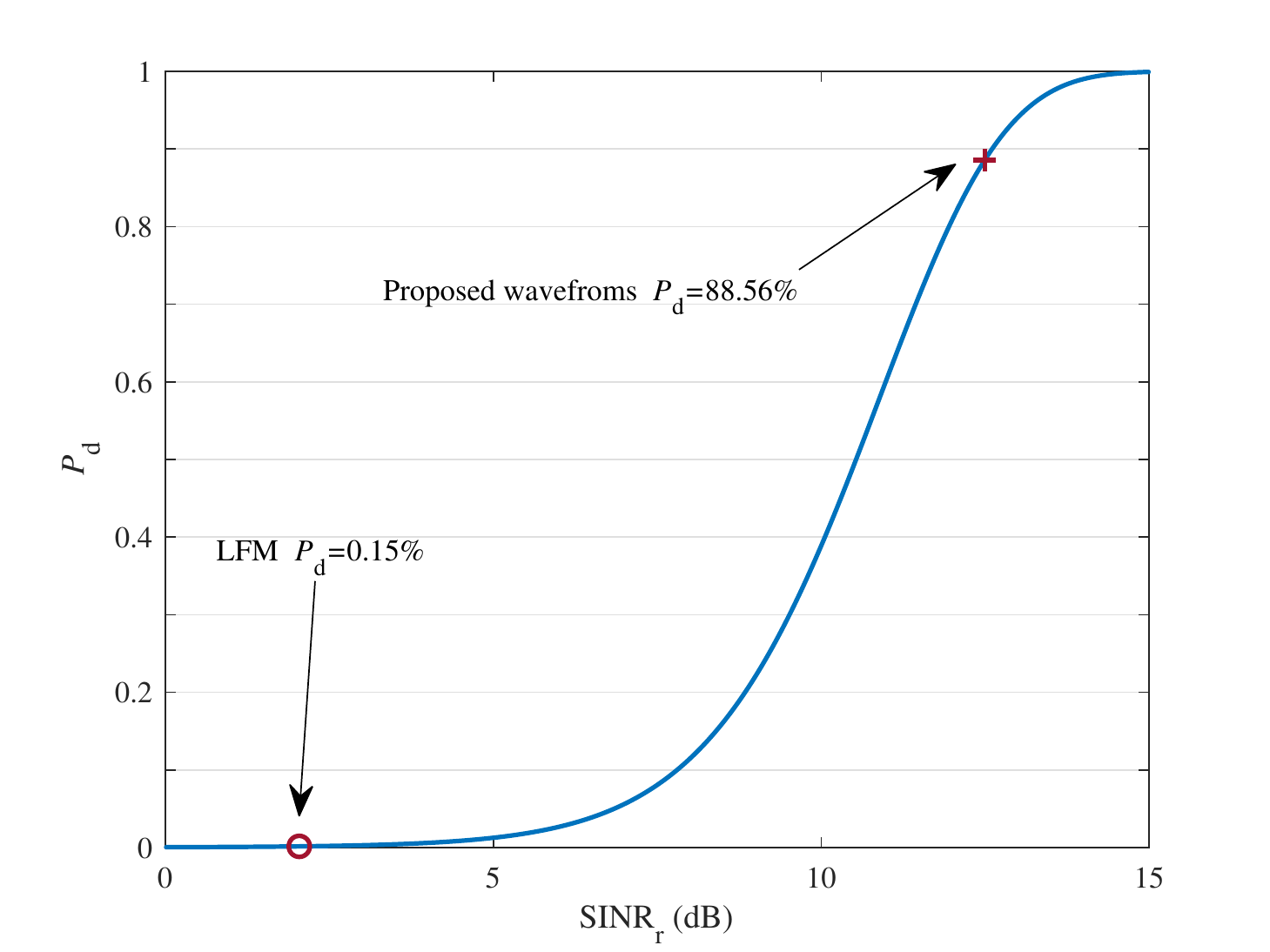}
	\caption{Target detection probability of the DFRC systems. $N_{\textrm{T}}=16,N_{\textrm{R}}=8,{e_{\textrm{T}}} = 20,L=20,\sigma_0^2 = -20$ dB.}
	\label{fig_4}
\end{figure*}

Next we compare the detection performance of the proposed waveforms with that of the LFM waveforms.  To detect the target, we set up a hypothesis test as follows:
\begin{equation}\label{eq:binary hypothesis}
	\left\{ \begin{array}{l}
		{{\cal H}_0: \by = \br},\\
		{{\cal H}_1: \by = \alpha_0 \bA (\theta_0) \bx +  \br},
	\end{array} \right.
\end{equation}
where $\br = \sum_{q = 1}^Q \alpha_q \bA (\theta_q) \bx +\bn$. Assume that $\bn \sim {\cal C} {\cal N}({\mathbf{0}},\sigma_{\textrm{n}}^2 \bI)$ and $\br \sim {\cal C} {\cal N}({\mathbf{0}},\bR_x)$ ($\bR_x$ is defined in \eqref{eq:transform 1}).
According to the Neyman-Pearson criterion \cite{kaybook1998}, we decide ${\cal H}_1$ if\footnote{\textcolor{red}{This detector is also called generalized matched filter \cite[pp. 478-479]{kaybook1998}}. We can also use the Bayesian detector proposed in \cite{Pd1973}, i.e., we decide ${\cal H}_1$ if $|\bw^\dagger \by|> T_{\text{h}}$.}
\begin{equation}\label{eq:H_1}
	\begin{split}
		\textrm{Re}(\bw^\dagger \by) > T_{\text{h}},
	\end{split}
\end{equation}
where $\bw = \alpha_0 \bR_x^{-1} \bA (\theta_0) \bx$, and $T_{\text{h}}$ is the detection threshold. The detection probability associated with this detector is given by\cite{kaybook1998} \footnote{The detection probability presented in \eqref{eq:P_D} represents an upper bound for the waveform $\bx$. If the prior knowledge of the clutter or the target is imprecise, the detection performance degrades. }
\begin{equation}\label{eq:P_D}
	\begin{split}
		P_{\textrm{D}} = \frac{1}{2}\textrm{erfc}\left(\textrm{erfc}^{-1}(2P_{\textrm{FA}})-\sqrt{\textrm{SINR}_{\rm{r}}}\right),
	\end{split}
\end{equation}
where $\textrm{erfc}(x)=\frac{2}{\sqrt{\pi}}\displaystyle\int_{x}^{\infty} \mathrm{e}^{-t^2}\mathrm{d}t$ is the complementary error function, $P_{\textrm{FA}}$ is the probability of false alarm, and $\textrm{SINR}_{\rm{r}}$ is given by
\begin{align}\label{eq:SINR new}
	\textrm{SINR}_{\rm{r}} = \sigma_0^2\bx^\dagger\bA^\dagger(\theta_0)\bR_x^{-1}\bA(\theta_0)\bx.
\end{align}
We can observe from \eqref{eq:P_D} and \eqref{eq:SINR new} that the detection probability $P_{\textrm{D}}$ depends on the transmit waveforms. Therefore, it can be expected that the optimized waveforms can achieve a larger SINR and a higher detection probability. To illustrate the superiority of the proposed waveforms, we let $\sigma_0^2 = -20$ dB. It can be verified that the $\textrm{SINR}_{\rm{r}}$ of the proposed waveform and the LFM signals are $12.49$ dB and $2.04$ dB, respectively. The associated detection probability of them is shown in \figurename~\ref{fig_4}. We can see that the detection probability of the proposed waveforms is significantly higher than that of the LFM signals.

\begin{figure*}[!htbp]
	\centering
	\includegraphics[width=2.9in]{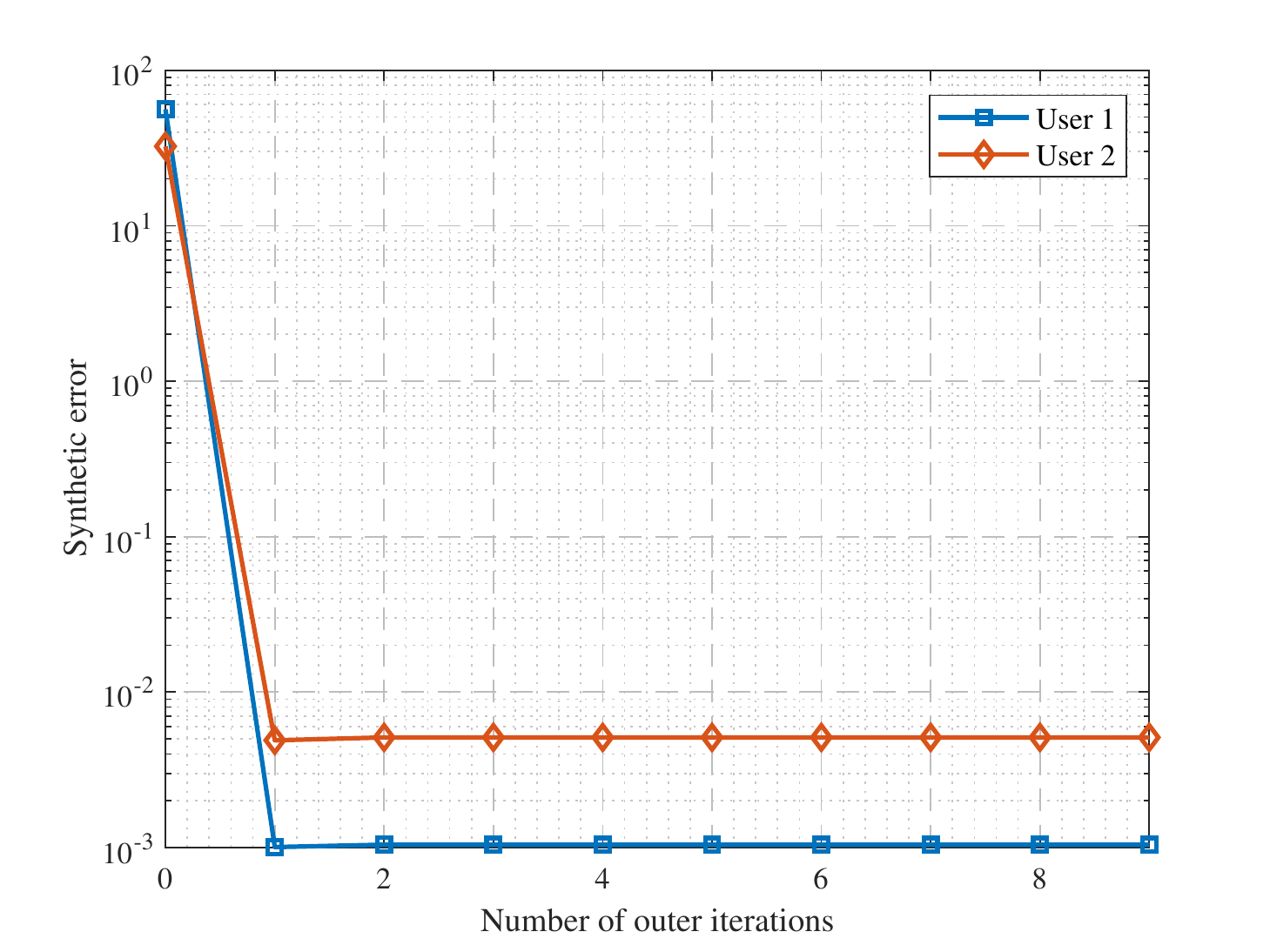}
	\caption{The synthesis error of communication signals versus the number of outer iterations. $M = 2, e_1 = e_2 = 20, L =20, \varsigma_1 = 10^{-3}, \varsigma_2 = 5\times 10^{-3}$.}
	\label{fig_5}
\end{figure*}

\begin{figure*}[!htbp]
\centering
{\subfigure[]{{\includegraphics[width = 2.1in]{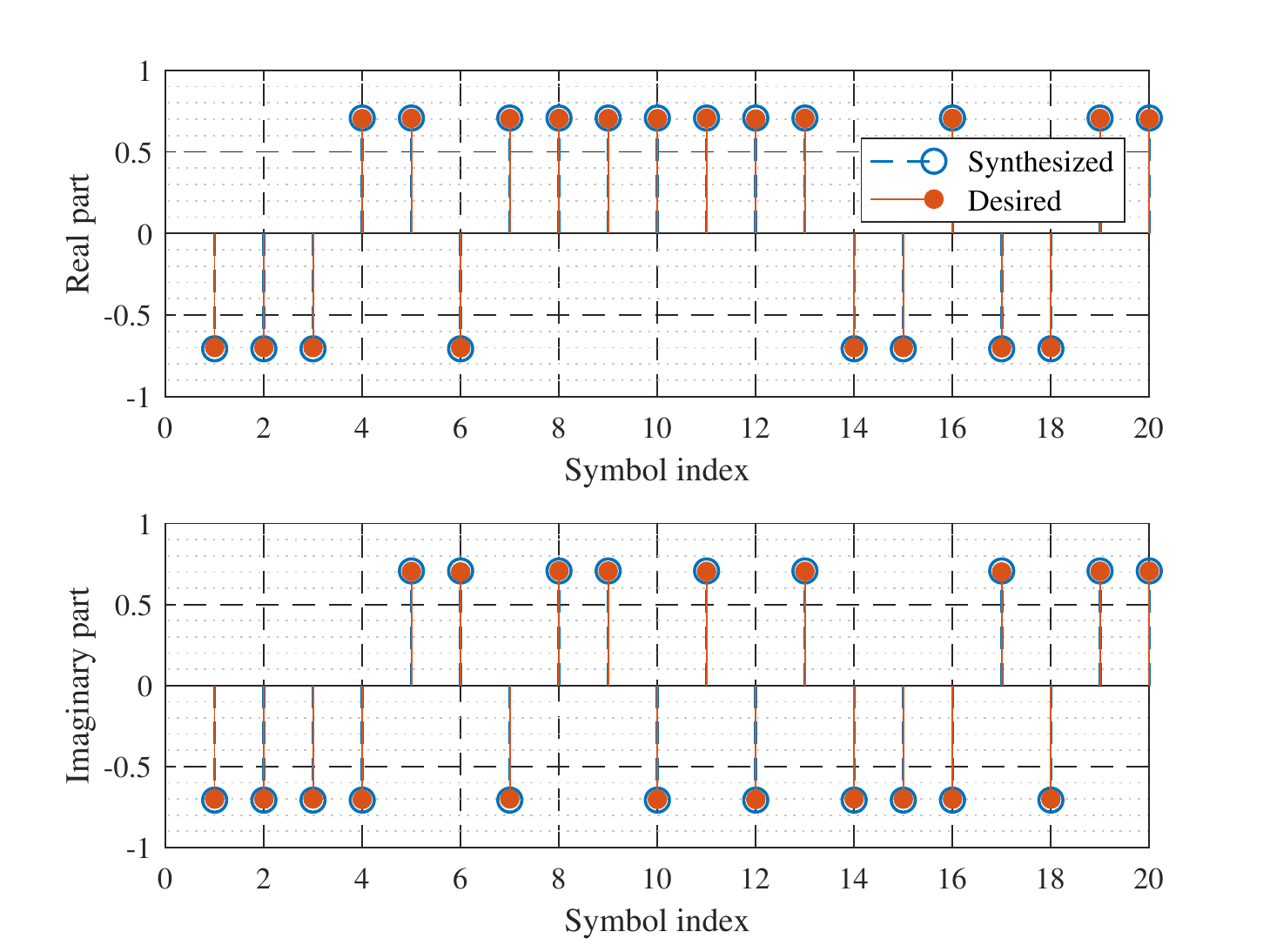}} \label{fig_6-1}} }
{\subfigure[]{{\includegraphics[width = 1.615in]{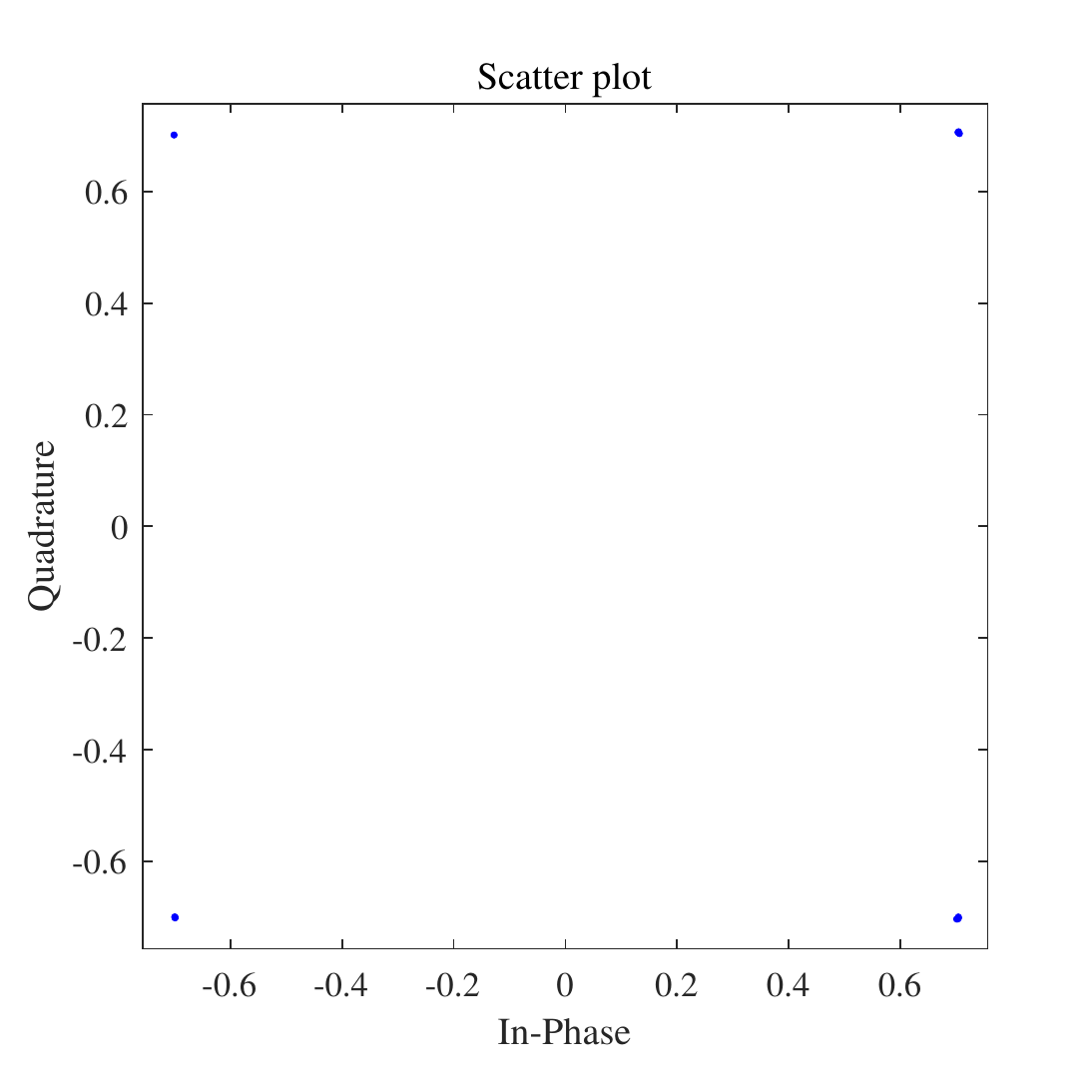}} \label{fig_6-2}} }
{\subfigure[]{{\includegraphics[width = 2.1in]{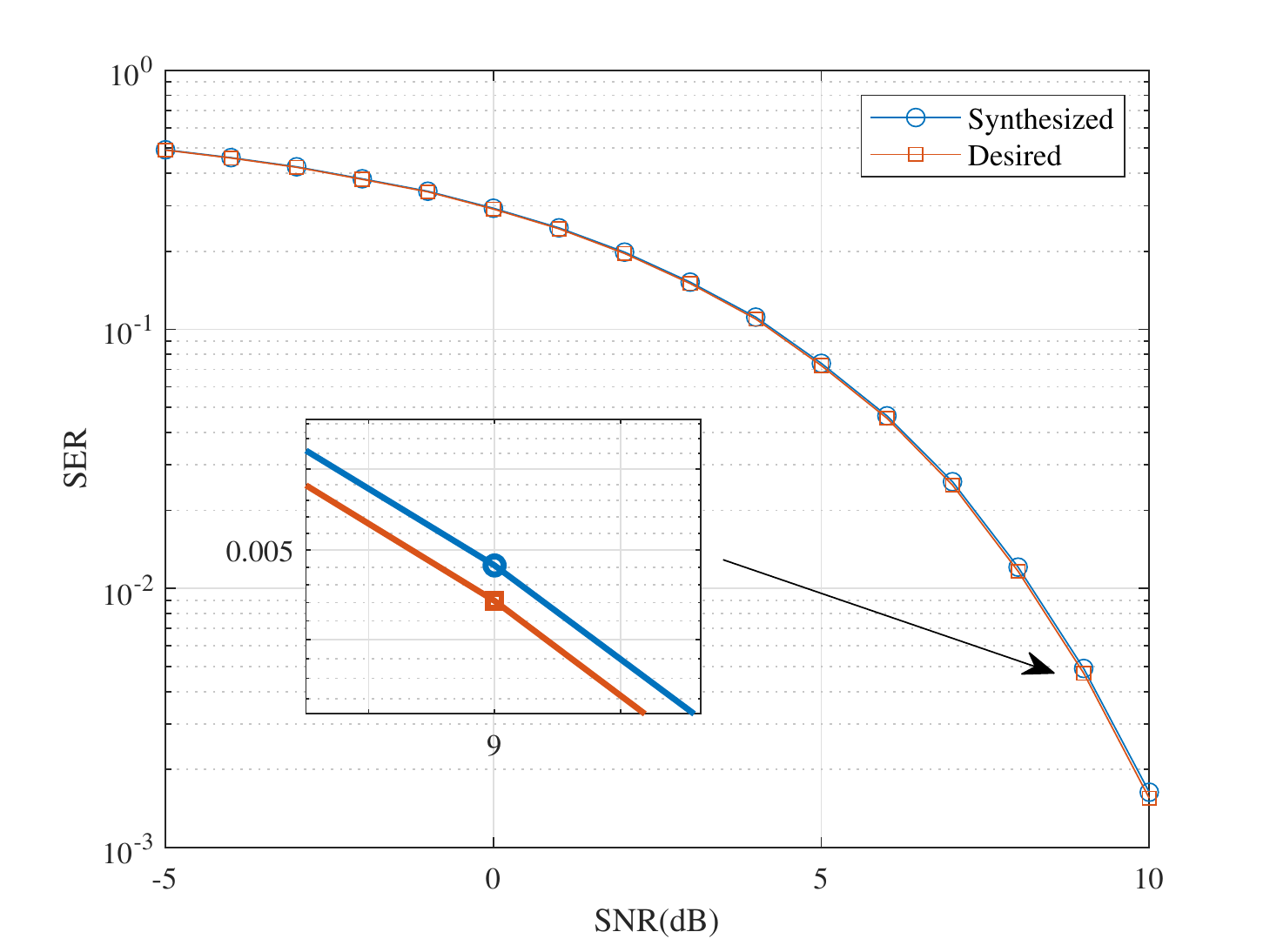}} \label{fig_6-3}} }
{\subfigure[]{{\includegraphics[width = 2.1in]{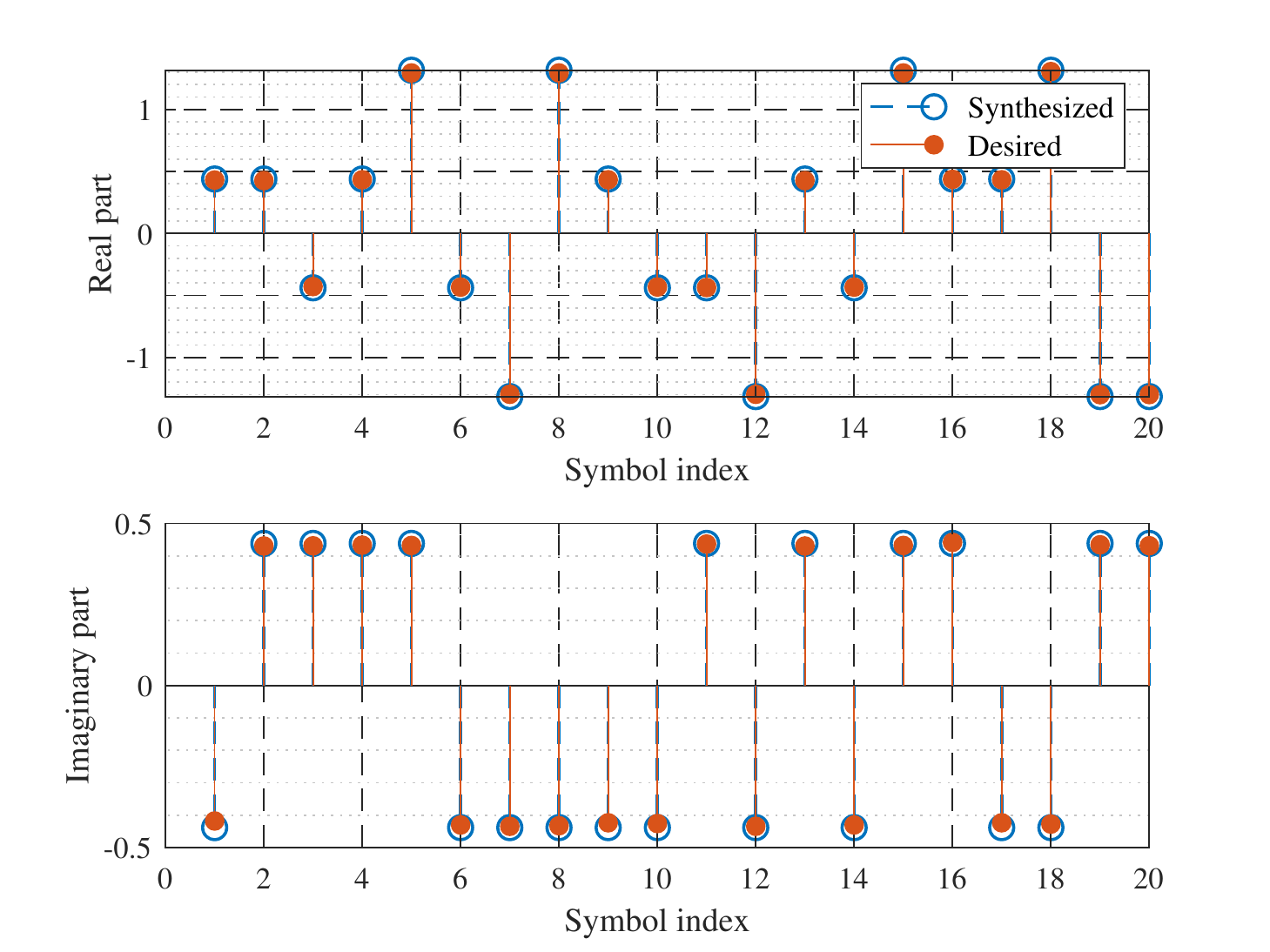}} \label{fig_6-4}} }
{\subfigure[]{{\includegraphics[width = 1.615in]{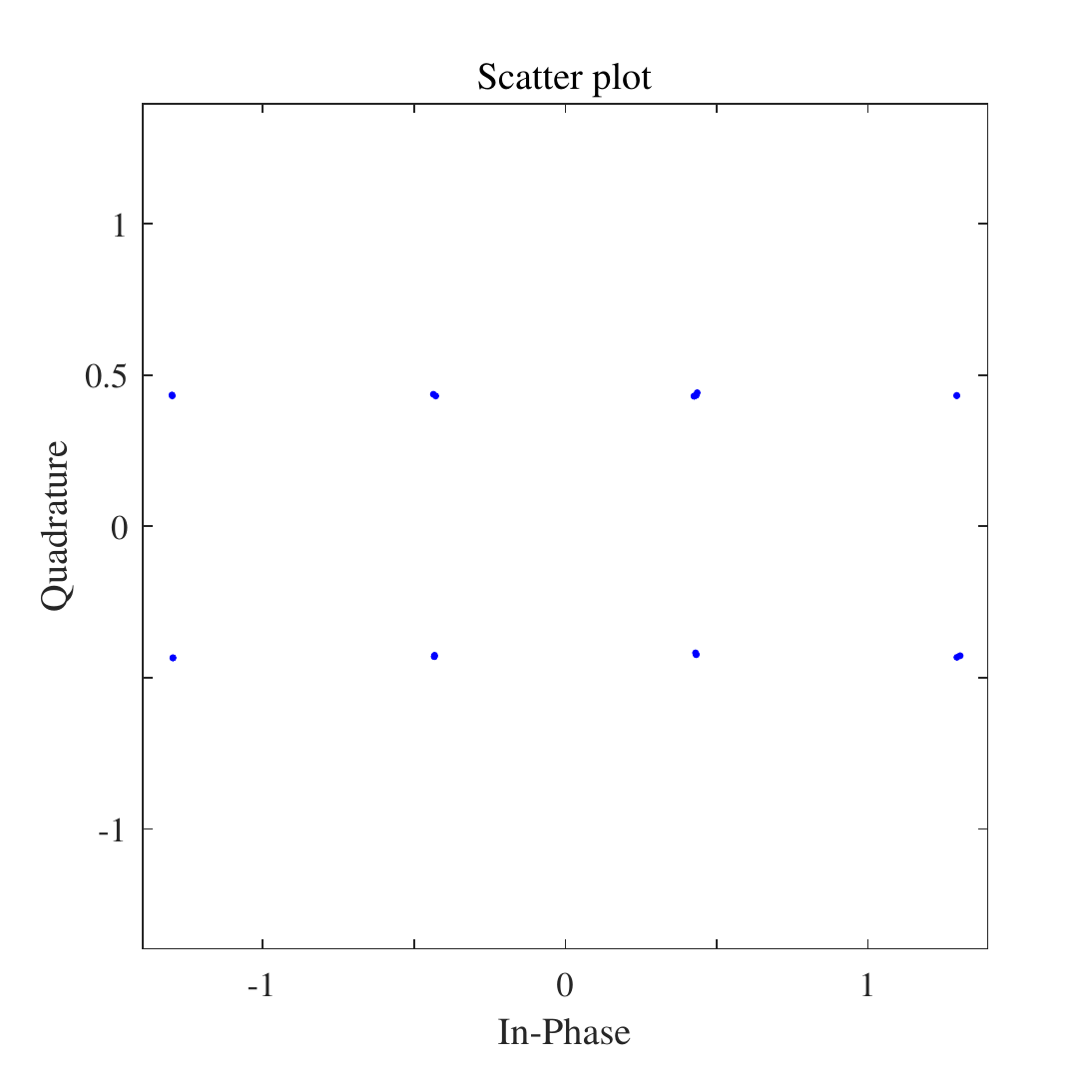}} \label{fig_6-5}} }
{\subfigure[]{{\includegraphics[width = 2.1in]{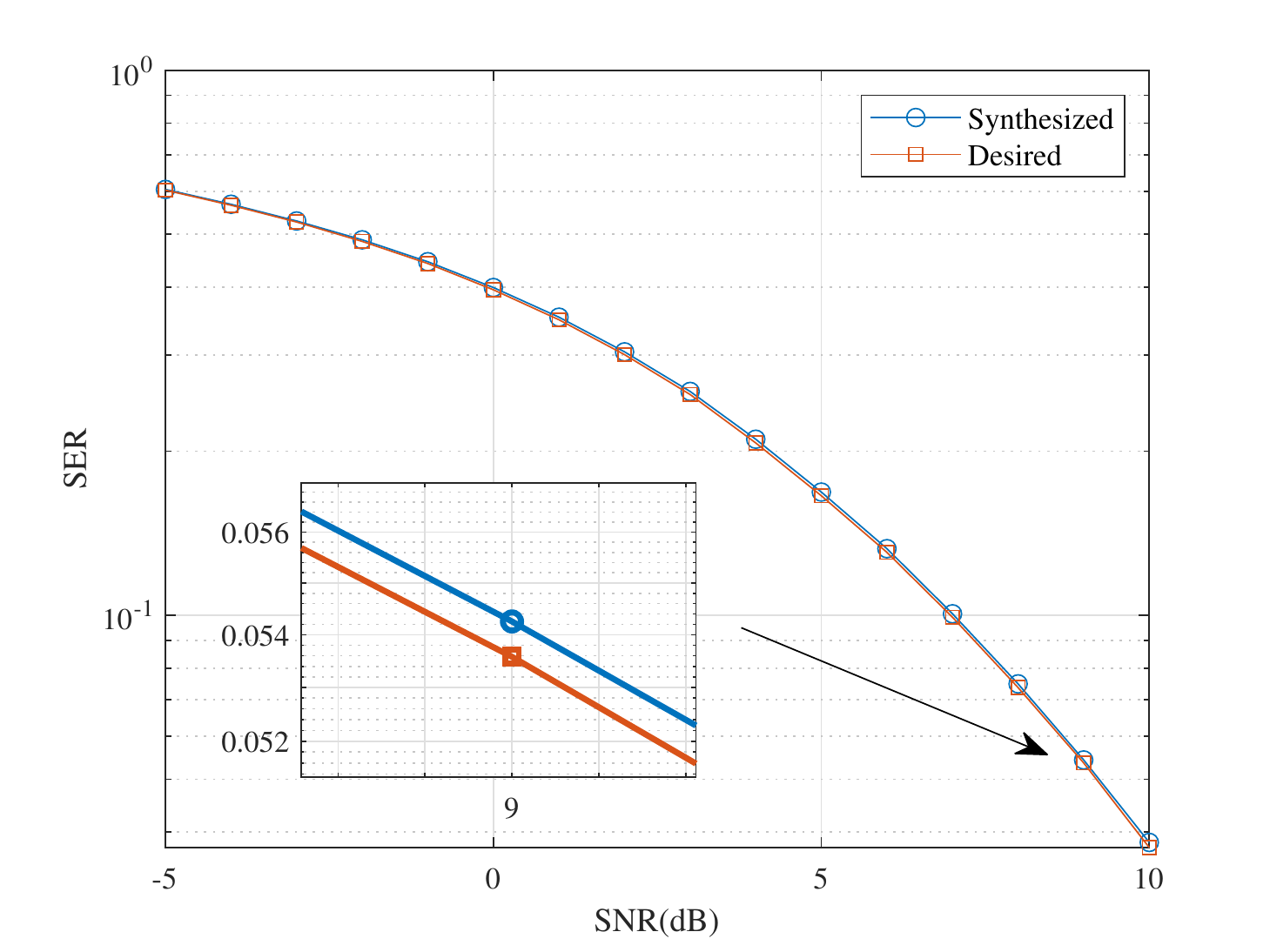}} \label{fig_6-6}} }
\caption{Analysis of the synthesized communication signals.  (a)-(c): Comparison of the synthesized communication signals with the desired ones, the constellation diagram, and the SER of the synthesized communication signals for user 1. (d)-(f): Comparison of the synthesized communication signals with the desired ones, the constellation diagram, and the SER of the synthesized communication signals for user 2. $M = 2, e_1 = e_2 = 20, L =20, \varsigma_1 = 10^{-3}, \varsigma_2 = 5\times 10^{-3}$.}
\label{fig_6}
\end{figure*}

\figurename~\ref{fig_5} shows the synthesis error of the communication signals versus the number of outer iterations. We can see that the waveforms at convergence satisfy the communication constraints, implying that the communication functionality is supported. To verify this claim, \figurename~\ref{fig_6}(a) and \figurename~\ref{fig_6}(d) compare the synthesized communication signals and the desired signals for the two users. The constellation diagrams of the synthesized communication signals are displayed in \figurename~\ref{fig_6}(b) and \figurename~\ref{fig_6}(e). It can be seen that the real and imaginary part of the synthesized signals are close to the desired ones and  the constellation diagrams of these signals are nearly ideal. \figurename~\ref{fig_6}(c) and \figurename~\ref{fig_6}(f) show the symbol error rate (SER) of the synthesized signals, where we define the SNR for the $m$th communication user as
\begin{equation}\label{eq:SNR}
	\begin{split}
		{\textrm{SNR}}_{m,c} = \frac{\expE\{ |s_{m,l}|^2\} }{\sigma_{z,m}^2},
	\end{split}
\end{equation}
$s_{m,l}$ is the $l$th symbol of $\bs_m$ ($l=1,2,\cdots,L$), $\sigma_{z,m}^2$ is the noise power level of the $m$th communication receiver, and $2000$ independent trials are conducted to obtain the SER. To achieve a given ${\textrm{SNR}}_{m,c}$, we keep the amplitude of the communication signals fixed and vary the noise power.
We can observe that the SER performance of the synthesized signals is also close to the desired ones.

\begin{figure*}[!htbp]
	\centering
	\includegraphics[width=2.9in]{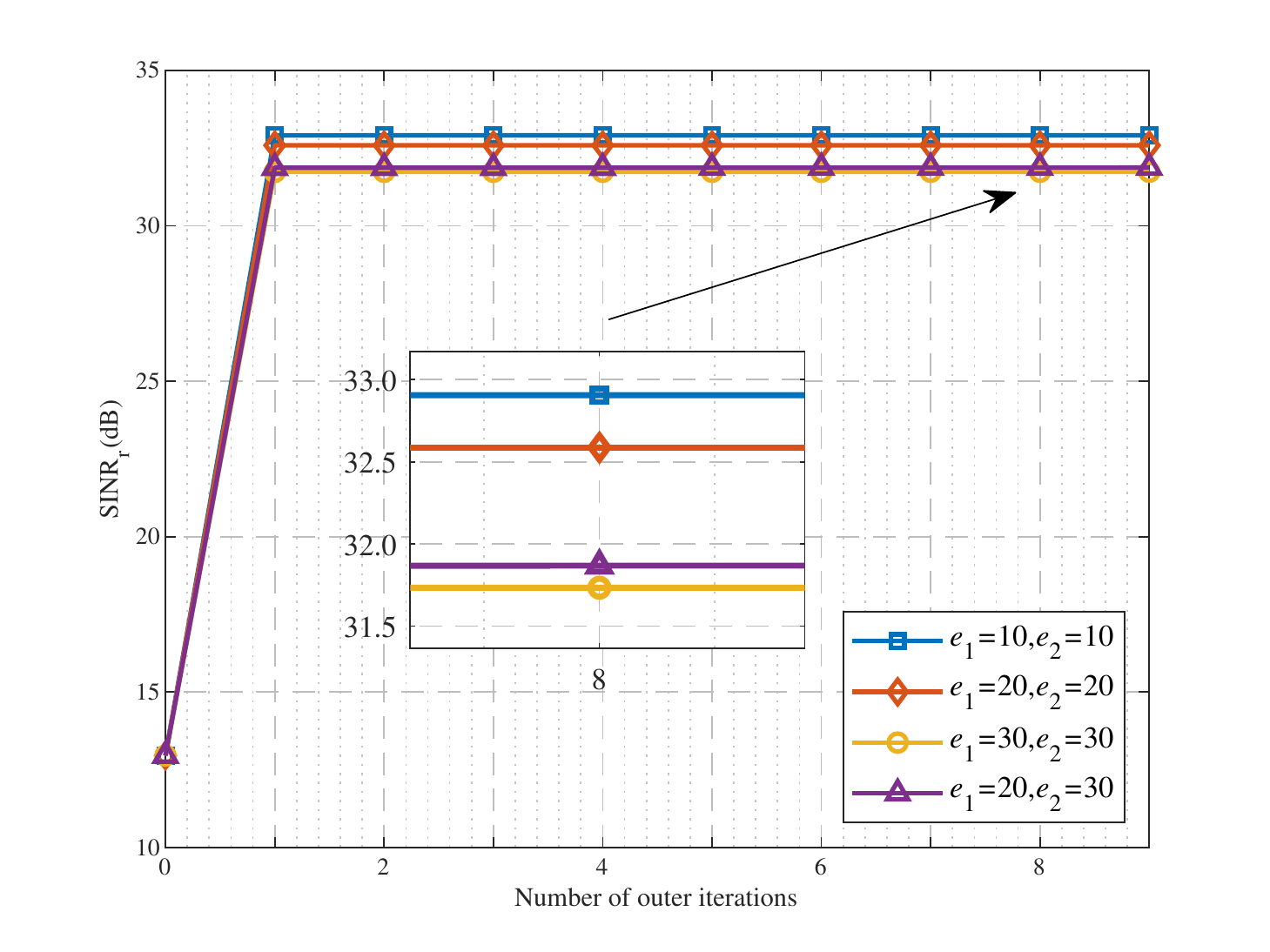}
	\caption{$\textrm{SINR}_{\rm{r}}$ for different $e_m$. $M = 2, L =20, \varsigma_1 = 10^{-3}, \varsigma_2 = 5\times 10^{-3}$.}
	\label{fig_7}
\end{figure*}

Next, we vary the transmit energy of the communication signals and analyze the $\textrm{SINR}$ performance for target detection. \figurename~\ref{fig_7} illustrates the $\textrm{SINR}$ curves of the DFRC system versus the number of outer iterations for different communication energy. The associated $\textrm{SINR}$ values at convergence are listed in Table \ref{tab:transmit energy}. Note that $\textrm{SINR}$ decreases with the transmit energy of the communication signals. This is due to that the DFRC system has to spare more transmit energy to ensure the communication performance, resulting in a degraded target detection performance.

\begin{table}[!htbp]
	\caption{$\textrm{SINR}_{\rm{r}}$ versus $e_m$}
	\centering
	\begin{tabular}{ccccc}
		\hline
        $e_1$  & $10$ & $20$ & $20$ & $30$ \\
        \hline
		$e_2$  & $10$ & $20$ &  $30$ & $30$ \\
		\hline
		$\textrm{SINR}_{\rm{r}}$ (dB) & $32.91$ & $32.59$ & $31.87$ & $31.73$\\
		\hline
	\end{tabular}
	\label{tab:transmit energy}
\end{table}

\begin{figure*}[!htbp]
	\centering
    {\subfigure[]{{\includegraphics[width = 2.5in]{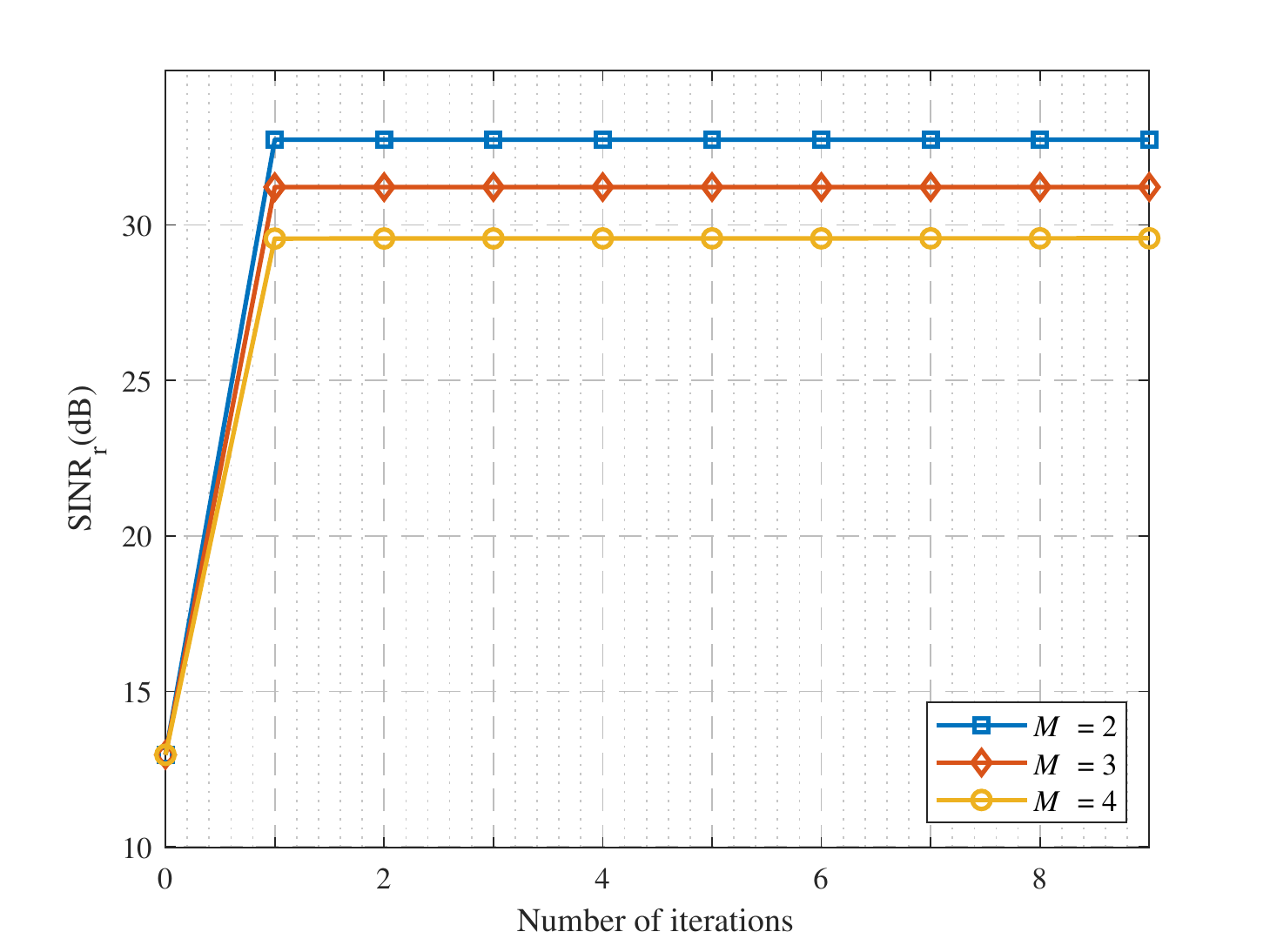}} \label{fig_8-1}} }
	{\subfigure[]{{\includegraphics[width = 2.5in]{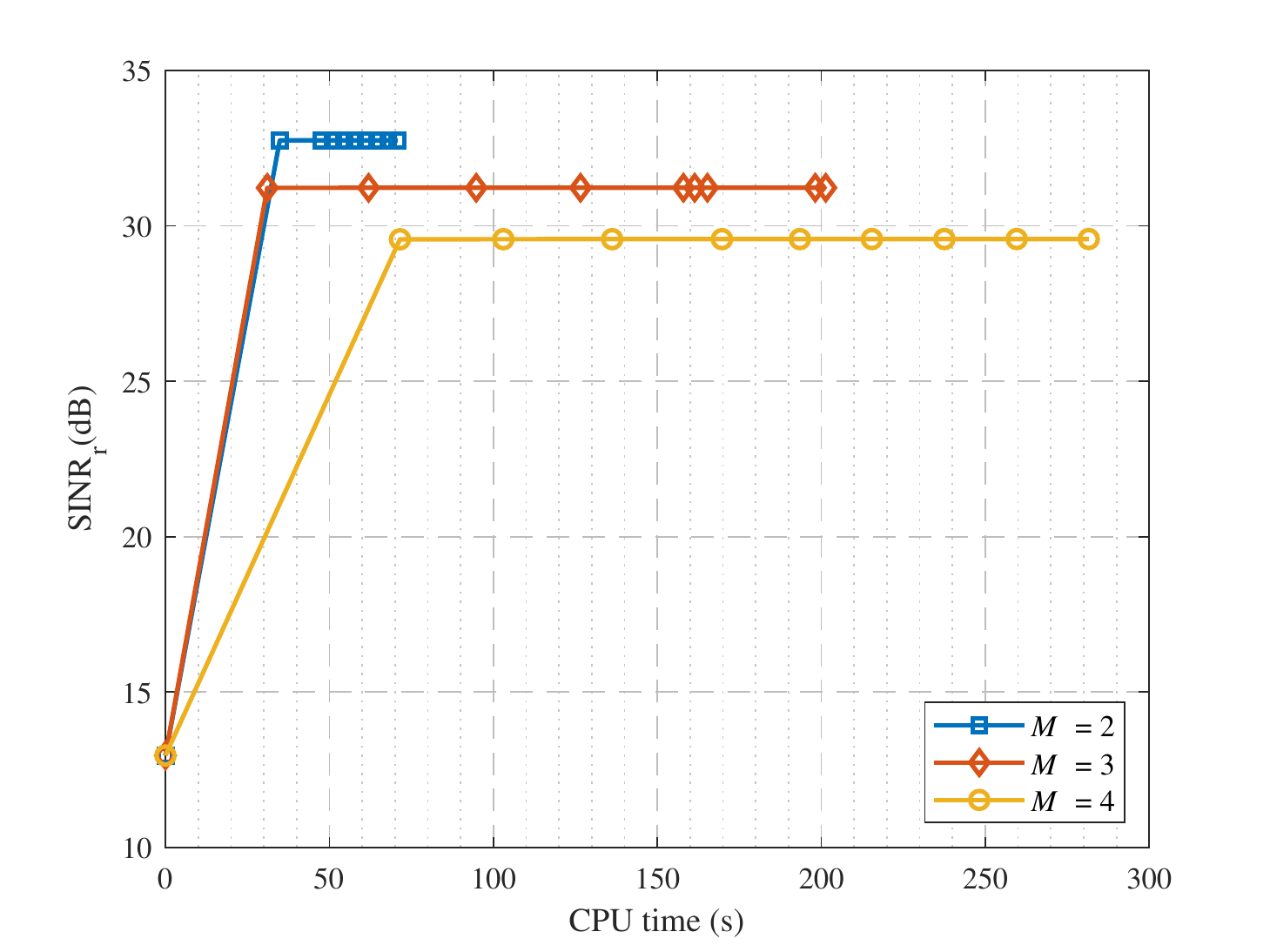}} \label{fig_8-2}} }
	\caption{$\textrm{SINR}_{\rm{r}}$ for different $M$. $L =20$, $e _m = 20, \varsigma_m =10^{-3}, m=1,\cdots,M$. (a) $\textrm{SINR}_{\rm{r}}$ versus the number of outer iterations. (b) $\textrm{SINR}_{\rm{r}}$ versus the CPU time.}
	\label{fig_8}
\end{figure*}

Then, we analyze the impact of the number of communication users on the system performance.
\figurename~\ref{fig_8} draws the $\textrm{SINR}$ curves versus the number of outer iterations for different number of communication users ($M=2,3$, and $4$), where the code length is $L = 20$, and the maximum allowed synthesis errors are $\varsigma_m =10^{-3}$ ($m=1, \cdots, M$). The $\textrm{SINR}$ values at convergence for different $M$ are shown in Table \ref{tab:different users}. We can find that due to the increase of the number of communication users (thus, a larger number of constraints and a smaller feasibility region), the $\textrm{SINR}$ performance decreases rapidly. Moreover, the propose algorithm takes a longer time to converge.

\begin{table}[!htbp]
\caption{$\textrm{SINR}_{\rm{r}}$ versus $M$}
\centering
\begin{tabular}{cccc}
	\hline
	$M$ & $2$ & $3$ & $4$ \\
	\hline
	$\textrm{SINR}_{\rm{r}}$ (dB) & $32.74$ & $31.22$ & $29.57$\\
	\hline
\end{tabular}
\label{tab:different users}
\end{table}

\begin{figure*}[!htbp]
	\centering
	{\subfigure[]{{\includegraphics[width = 2.5in]{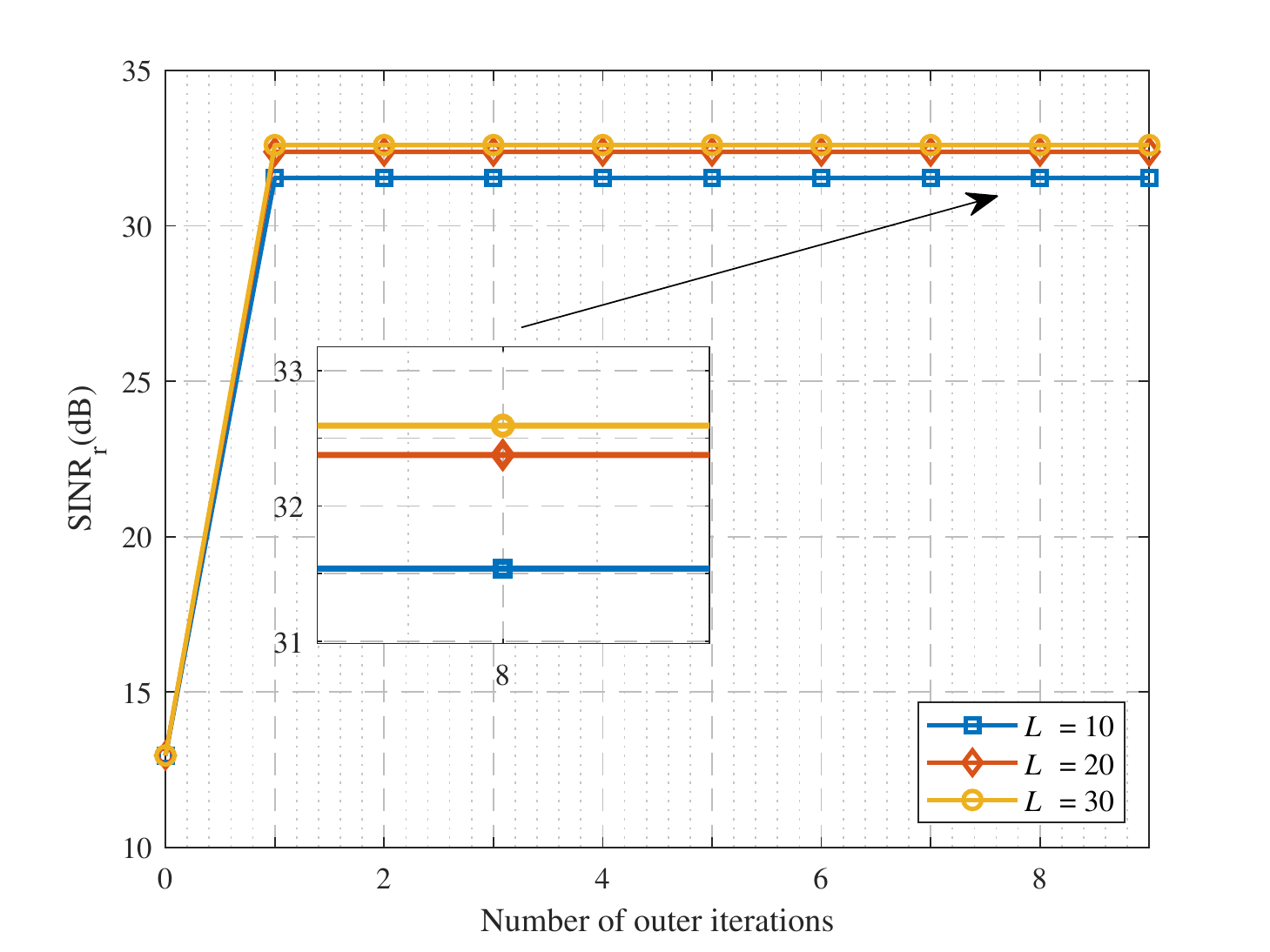}} \label{fig_9-1}} }
	{\subfigure[]{{\includegraphics[width = 2.5in]{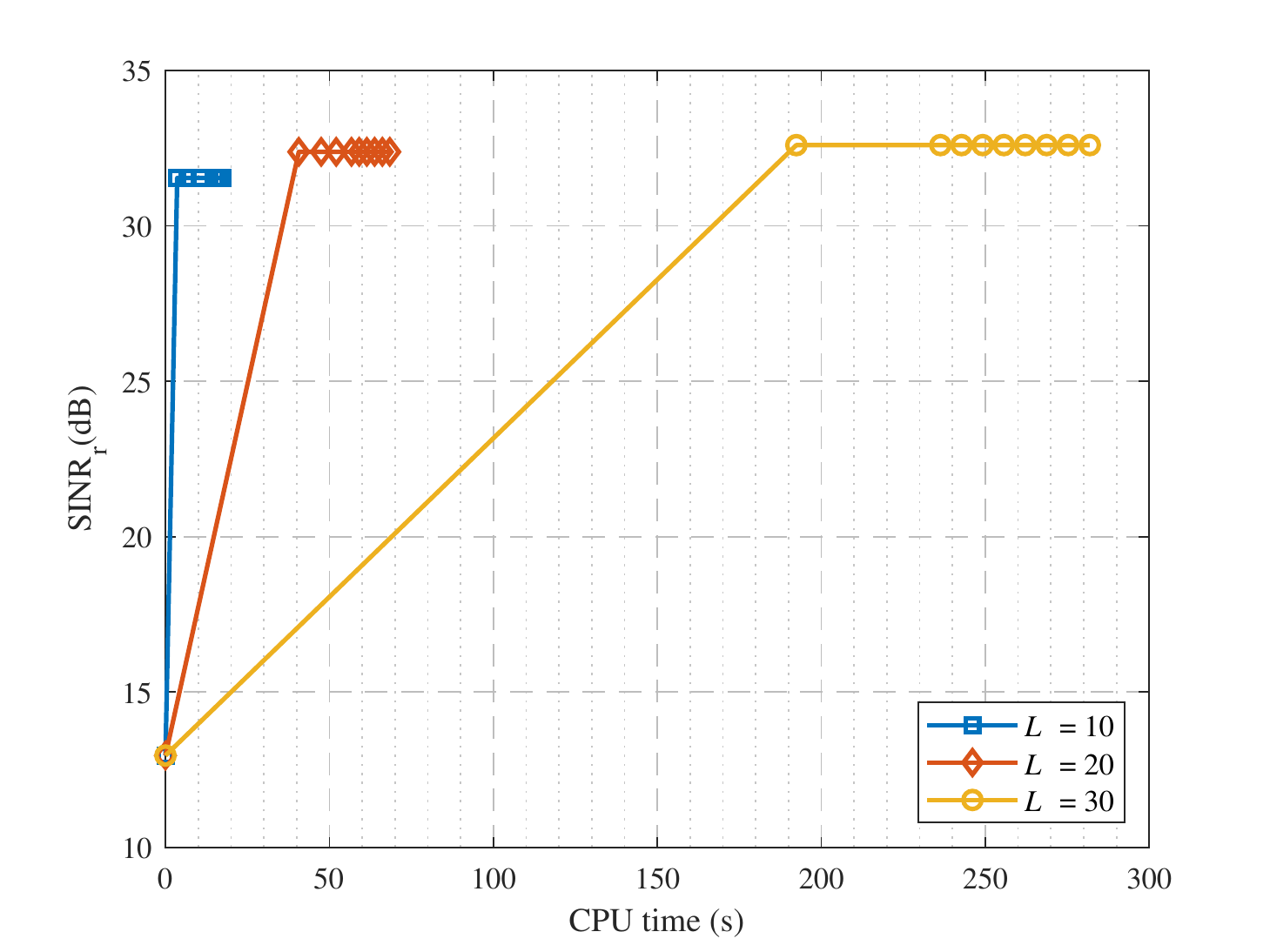}} \label{fig_9-2}} }
	\caption{$\textrm{SINR}_{\rm{r}}$ for different $L$. $M=2, e_1 = e_2 = 20, \varsigma_1 = \varsigma_2 = 10^{-3}$. (a) $\textrm{SINR}_{\rm{r}}$ versus the number of outer iterations. (b) $\textrm{SINR}_{\rm{r}}$ versus the CPU time.}
	\label{fig_9}
\end{figure*}

Now we study how the code length affects the DFRC system performance. We use the same parameter setting as that in \figurename ~\ref{fig_2}, but vary the code length and set the maximum allowed synthesis error to be $\varsigma_m =10^{-3}$ ($m=1,2$). \figurename~\ref{fig_9-1} and \figurename~\ref{fig_9-2} plot the $\textrm{SINR}$ of the synthesized waveforms versus the number of outer iterations and the CPU time, respectively, for $L = 10,20$, and $30$. Table \ref{tab:code length} presents the $\textrm{SINR}$ values at convergence. \figurename~\ref{fig_9-1} indicates that the target detection performance improves for a larger $L$. It is because that the degree of freedom (DOF) of the waveforms increases with the code length (which implies that the system has better interference suppression capability). However, as shown in \figurename~\ref{fig_9-2}, the increased code length will also lead to a longer time for the algorithm to reach convergence.

\begin{table}[!htbp]
\caption{$\textrm{SINR}_{\rm{r}}$ versus $L$}
\centering
\begin{tabular}{cccc}
	\hline
	$L$ & $10$ & $20$ & $30$\\
	\hline
	$\textrm{SINR}_{\rm{r}}$ (dB) & $31.54$ & $32.38$ & $32.59$\\
	\hline
\end{tabular}
\label{tab:code length}
\end{table}

\begin{figure*}[!htbp]
	\centering
	{\subfigure[]{{\includegraphics[width = 2.5in]{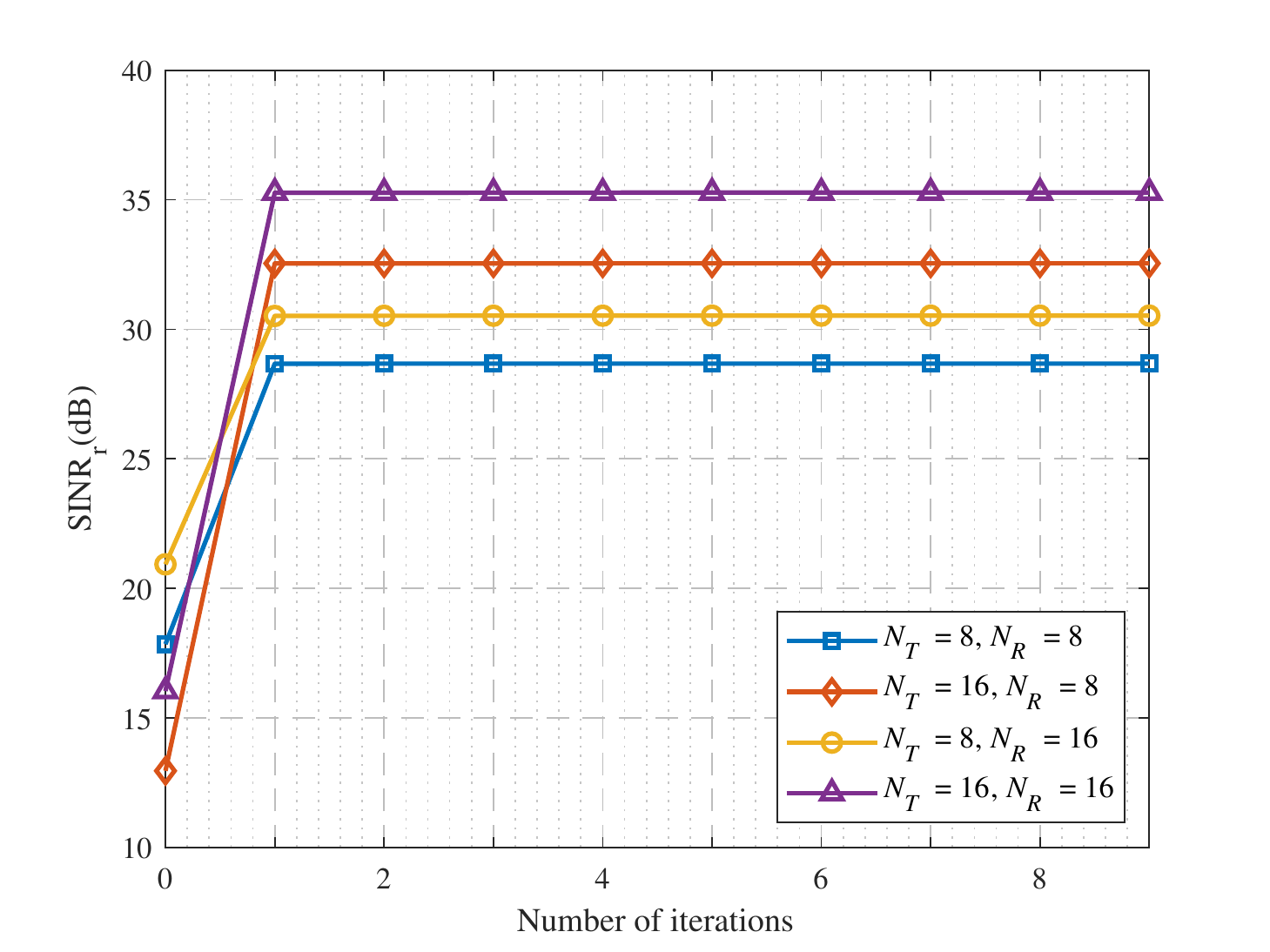}} \label{fig_10-1}} }
	{\subfigure[]{{\includegraphics[width = 2.5in]{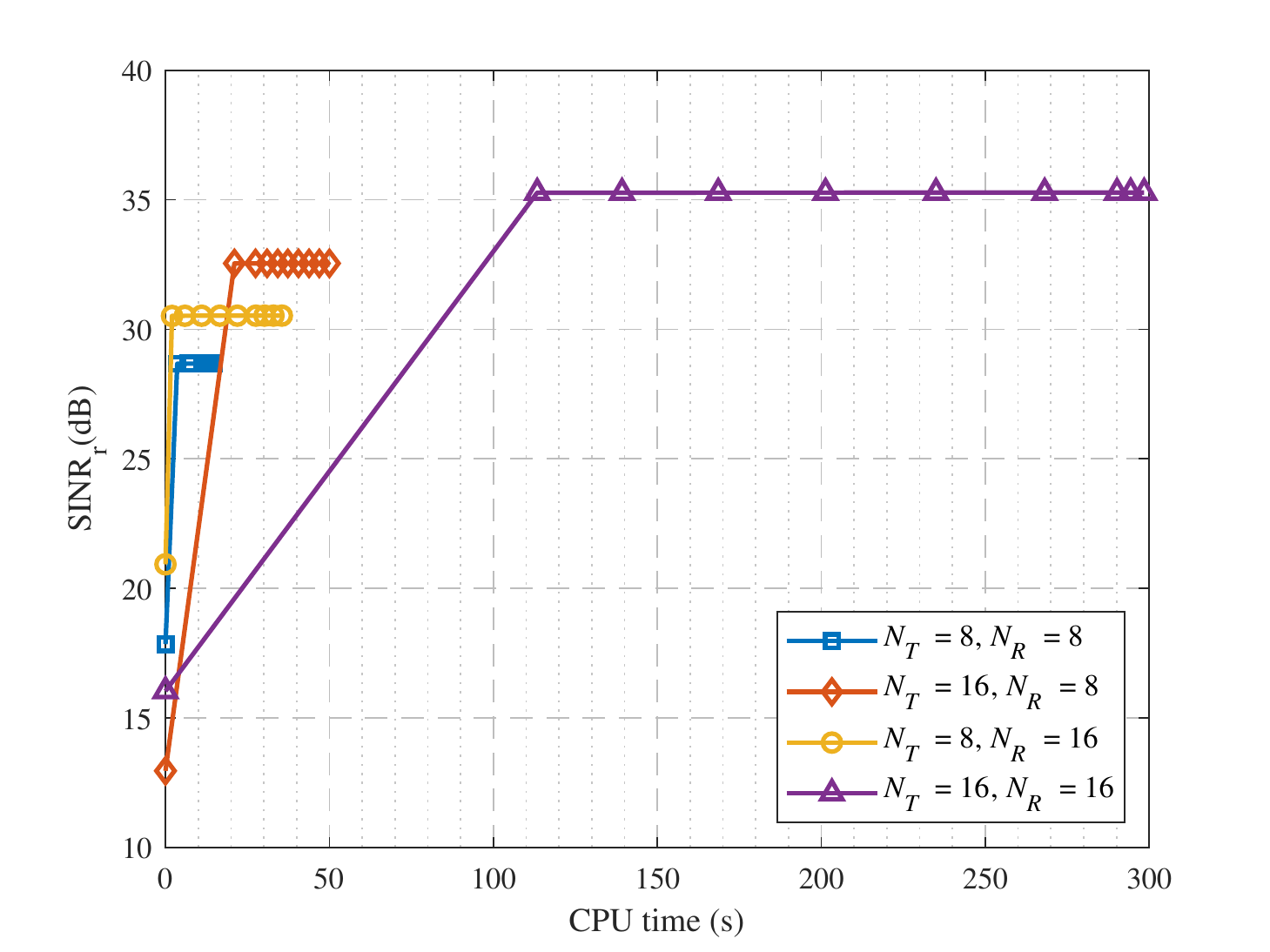}} \label{fig_10-2}} }
	\caption{$\textrm{SINR}_{\rm{r}}$ for different $N_\textrm{T}, N_\textrm{R}$. $L=20, M=2, e_1 = e_2 = 20, \varsigma_1 = \varsigma_2 = 10^{-3}$. (a) $\textrm{SINR}_{\rm{r}}$ versus the number of outer iterations. (b) $\textrm{SINR}_{\rm{r}}$ versus the CPU time.}
	\label{fig_10}
\end{figure*}

Subsequently, we study the impact of the number of antennas on the DFRC system performance. We use the same parameter setting as that in \figurename~\ref{fig_2}, but vary the number of antennas and set the maximum allowed synthesis error to be $\varsigma_m =10^{-3}$ ($m=1,2$). \figurename~\ref{fig_10}(a) and \figurename~\ref{fig_10}(b) show the convergence of the proposed algorithm versus the number of outer iterations and the CPU time, respectively, for different number of transmit and receive antennas. The $\textrm{SINR}$ values at convergence are displayed in Table \ref{tab:the number of antennas}. It can be seen that increasing the number of antennas improves the detection performance, but it also lead to a heavier computation load.

\begin{table}[!htbp]
	\caption{$\textrm{SINR}_{\rm{r}}$ versus $N_\textrm{T}$ and $N_\textrm{R}$}
	\centering
	\begin{tabular}{ccccc}
		\hline
		$N_\textrm{T}$ & $8$ & $8$ & $16$ & $16$\\
	    \hline
	    $N_\textrm{R}$ & $8$ & $16$ & $8$ & $16$\\
		\hline
		$\textrm{SINR}_{\rm{r}}$ (dB) & $28.68$ & $30.53$ & $32.55$ & $35.28$\\
		\hline
	\end{tabular}
	\label{tab:the number of antennas}
\end{table}

\begin{figure*}[!htbp]
\centering
\includegraphics[width=3in]{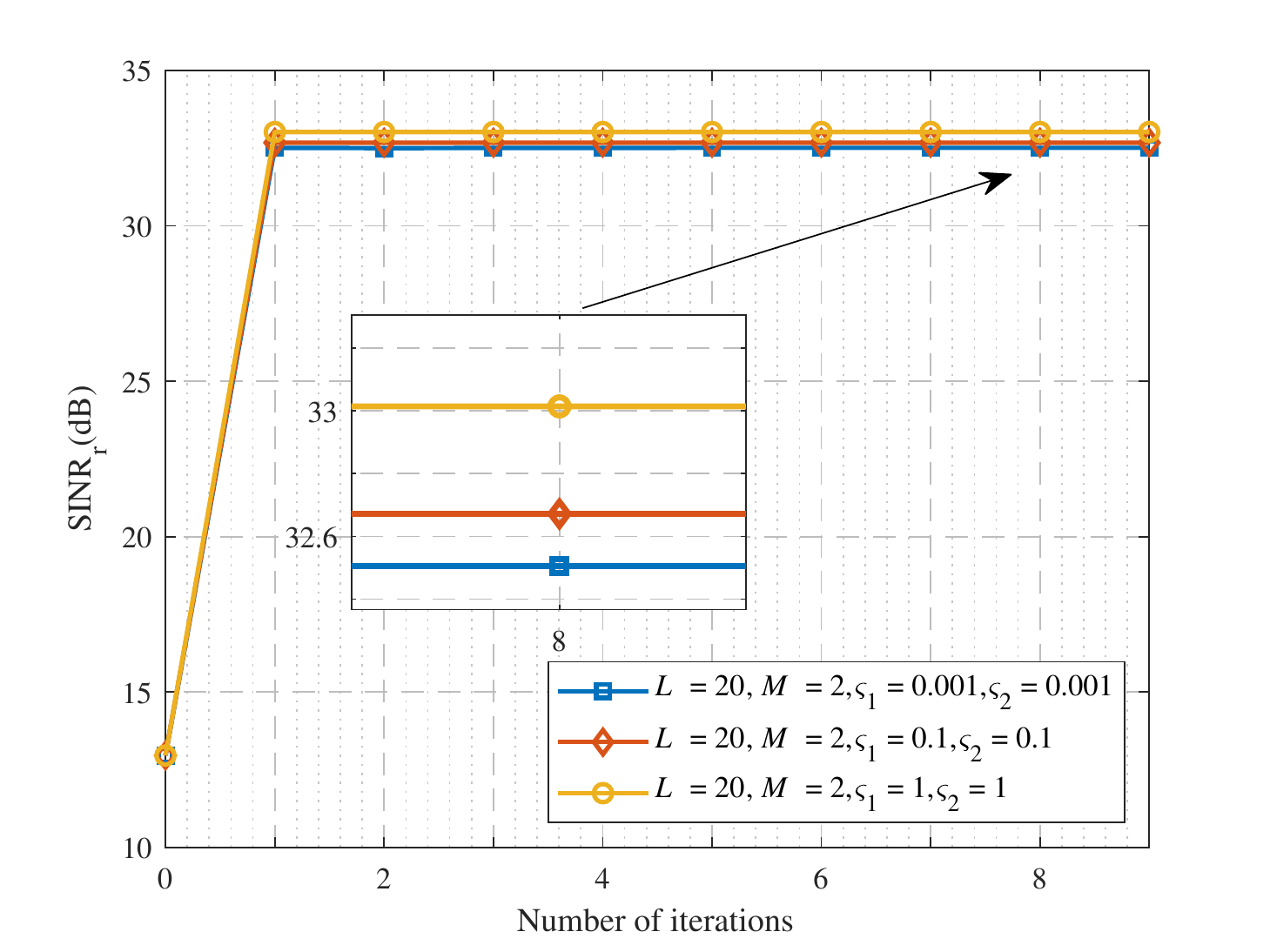}
\caption{$\textrm{SINR}_{\rm{r}}$ for different $\varsigma_m$. $M = 2, e_1 = e_2 = 20, L =20, \varsigma_1 = \varsigma_2$.}
\label{fig_11}
\end{figure*}

\begin{figure*}[!htbp]
	\centering
	{\subfigure[]{{\includegraphics[width = 2.5in]{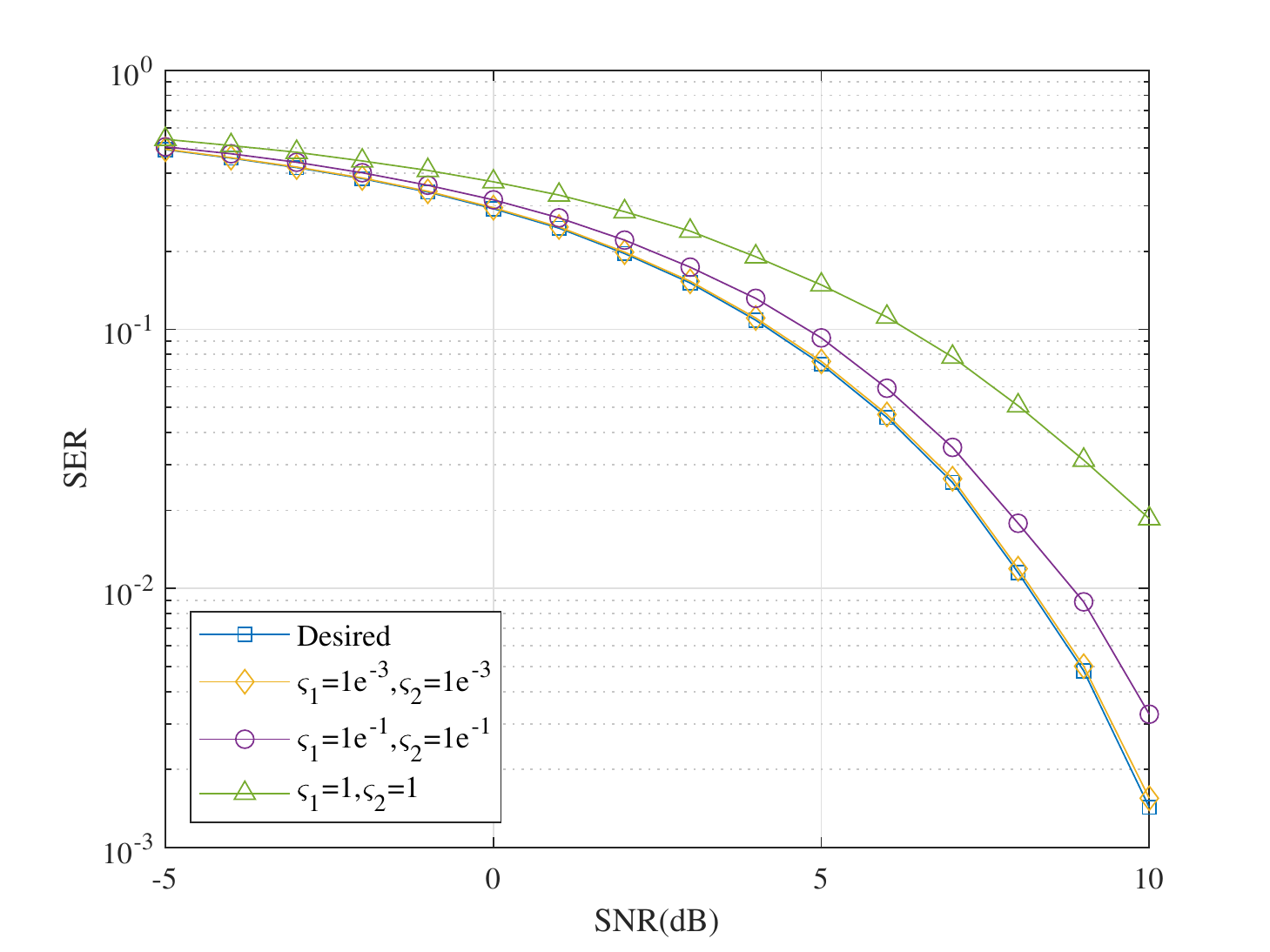}} \label{fig_12-1}} }
	{\subfigure[]{{\includegraphics[width = 2.5in]{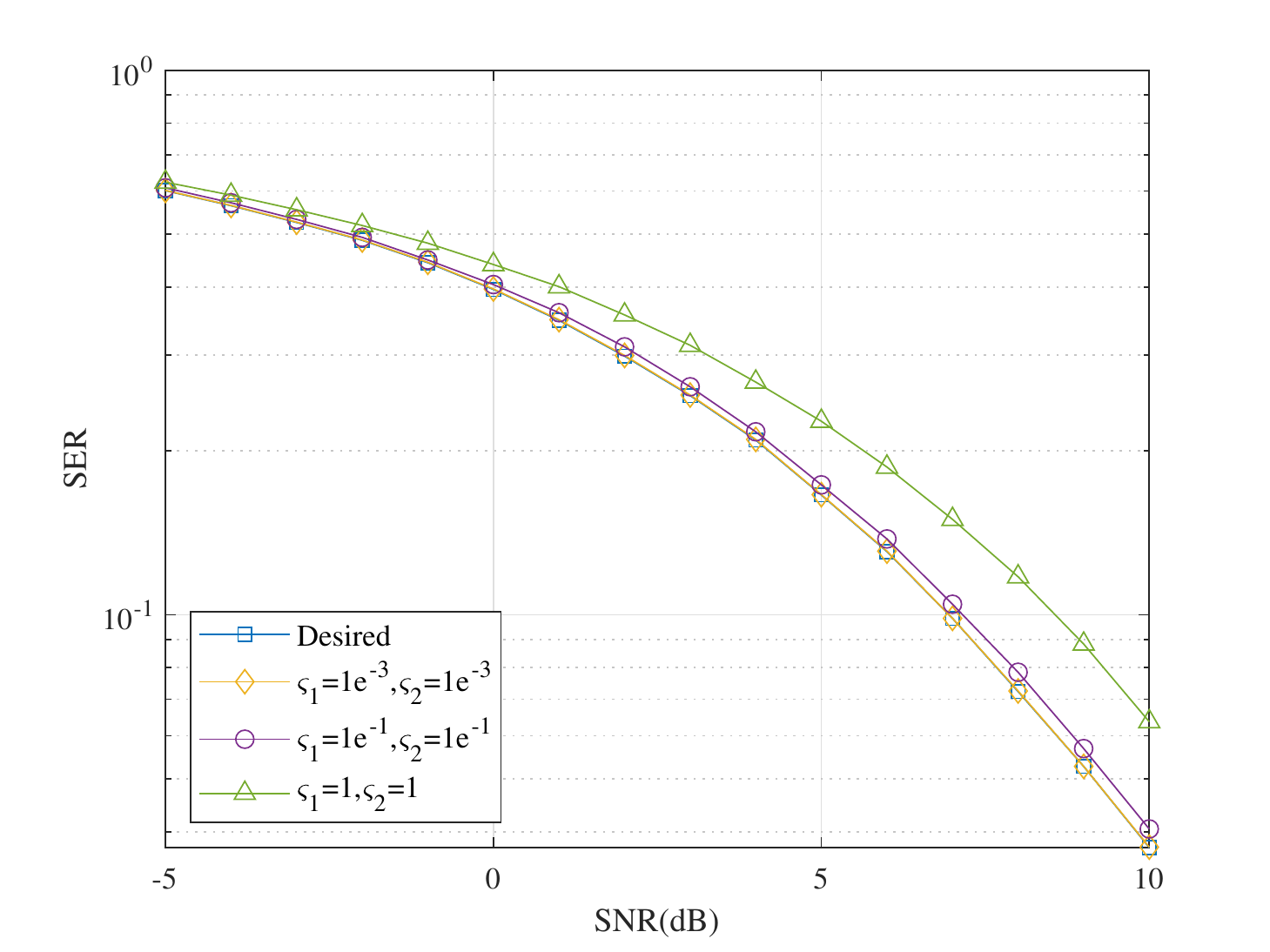}} \label{fig_12-2}} }
	\caption{SER for different $\varsigma_m$. $M=2, e_1 = e_2 = 20, L =20, \varsigma_1 = \varsigma_2$. (a) User 1. (b) User 2.}
	\label{fig_12}
\end{figure*}

In \figurename~\ref{fig_11}, the impact of the maximum allowed synthesis error on the $\textrm{SINR}$ performance is analyzed, where we use the same parameter setting as that in \figurename~\ref{fig_2} except for varying the maximum allowed synthesis error. The $\textrm{SINR}$ values at convergence for these cases are given in Table \ref{tab:the different synthesis error}. The SER performance of the synthesized communication signals is analyzed in \figurename~\ref{fig_12}. We can see that for a smaller value of the maximum allowed synthesis error, the $\textrm{SINR}$ performance only degrades slightly. However, the associated communication performance improves quickly. This implies that to guarantee the quality of service for communication without affecting the detection performance, we can control the synthesis error to a reasonably low level.

\begin{table}[!htbp]
\caption{$\textrm{SINR}_{\rm{r}}$ versus $\varsigma_m$}
\centering
\begin{tabular}{cccc}
	\hline
	$\varsigma_m$ & $10^{-3}$ & $10^{-1}$ & $1$ \\
	\hline
	$\textrm{SINR}_{\rm{r}}$ (dB) & $32.50$ & $32.67$ & $33.01$\\
	\hline
\end{tabular}
\label{tab:the different synthesis error}
\end{table}

\begin{figure*}[!htbp]
	\centering
	\includegraphics[width=3in]{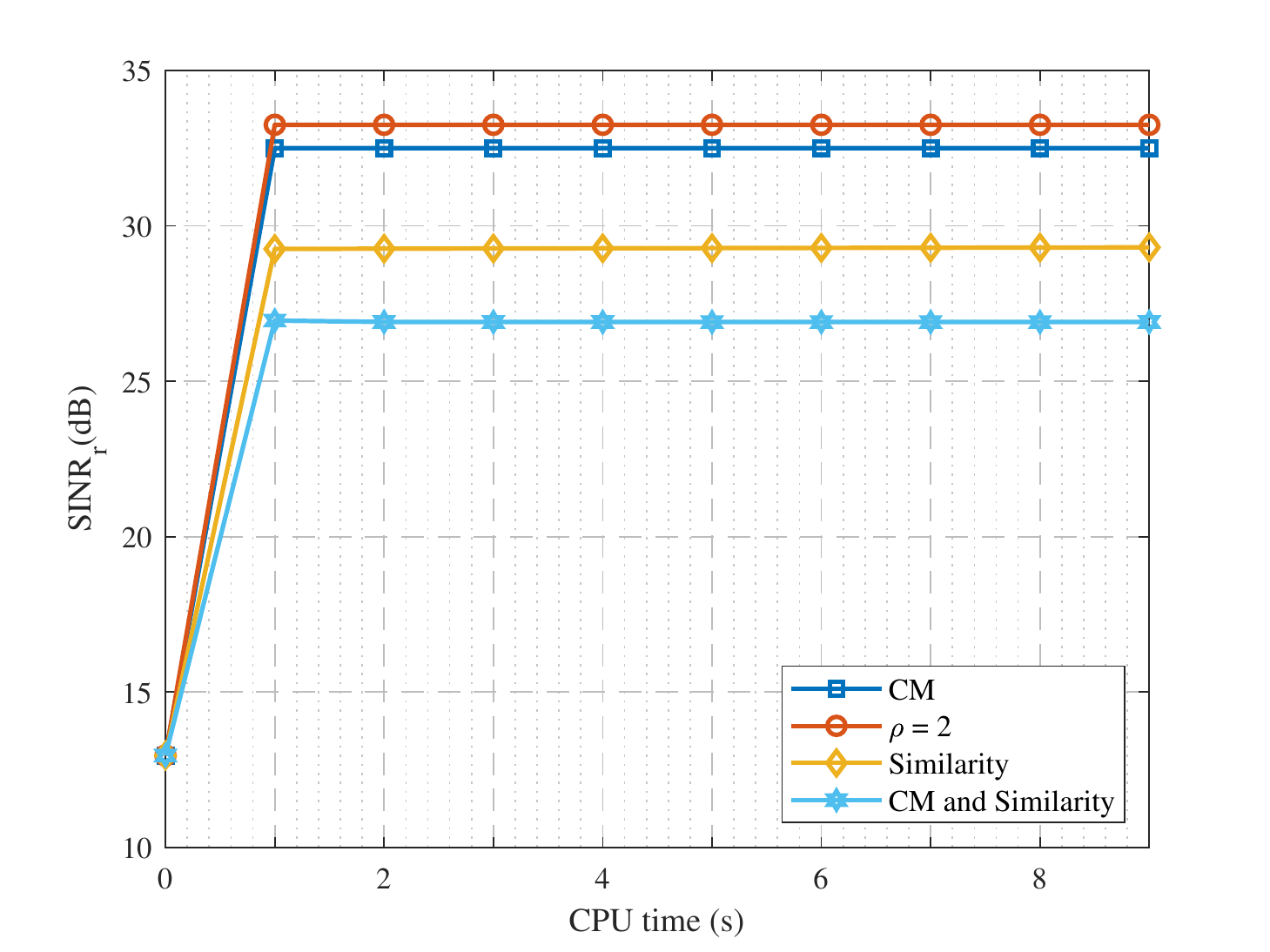}
	\caption{$\textrm{SINR}_{\rm{r}}$ under different constraints. $M = 2, e_1 = e_2 = 20, L =20, \varsigma_1 = 10^{-3}, \varsigma_2 = 5\times 10^{-3}$.}
	\label{fig_13}
\end{figure*}

\begin{figure*}[!htbp]
	\centering
	{\subfigure[]{{\includegraphics[width = 2.5in]{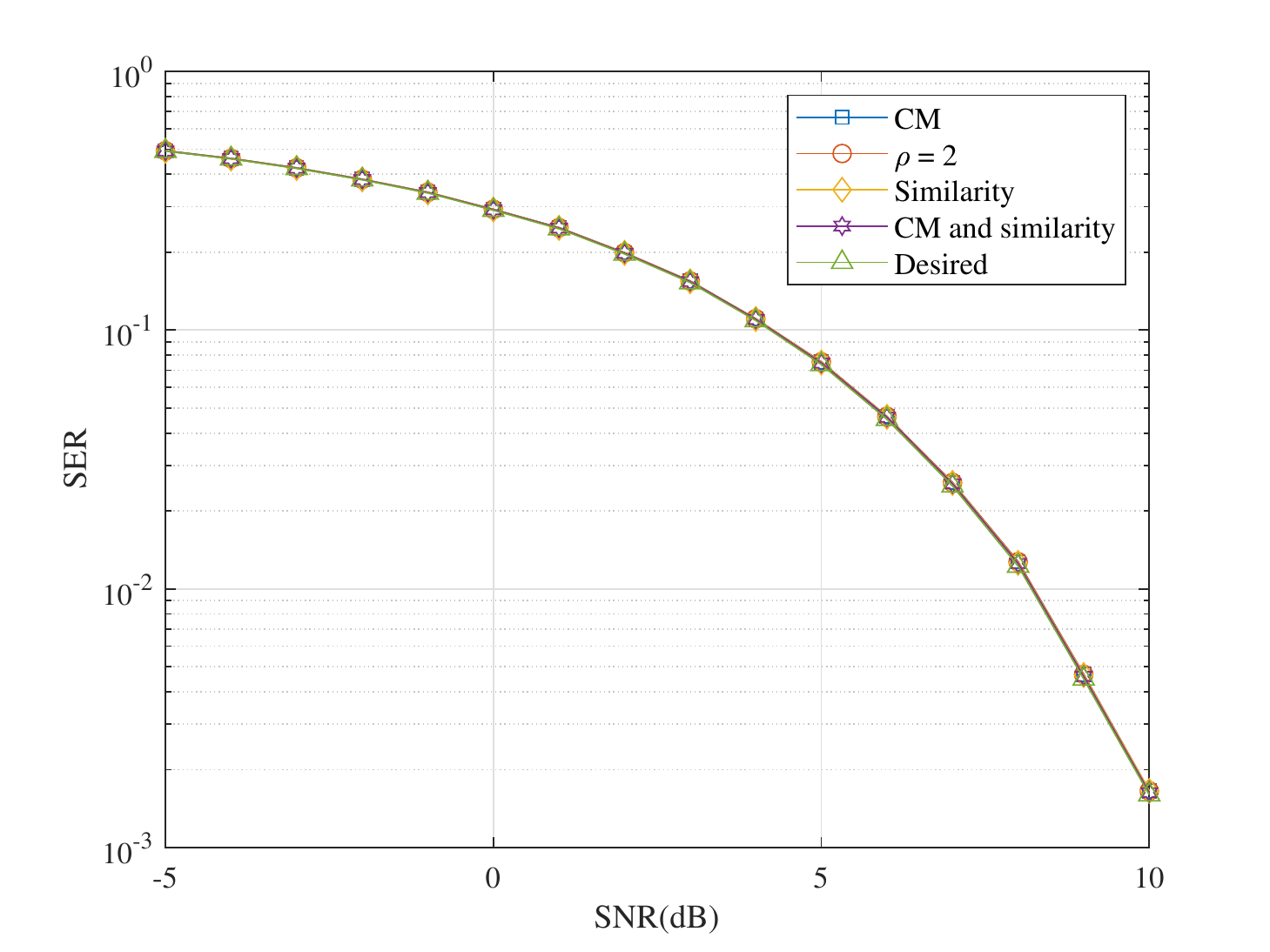}} \label{fig_14-1}} }
	{\subfigure[]{{\includegraphics[width = 2.5in]{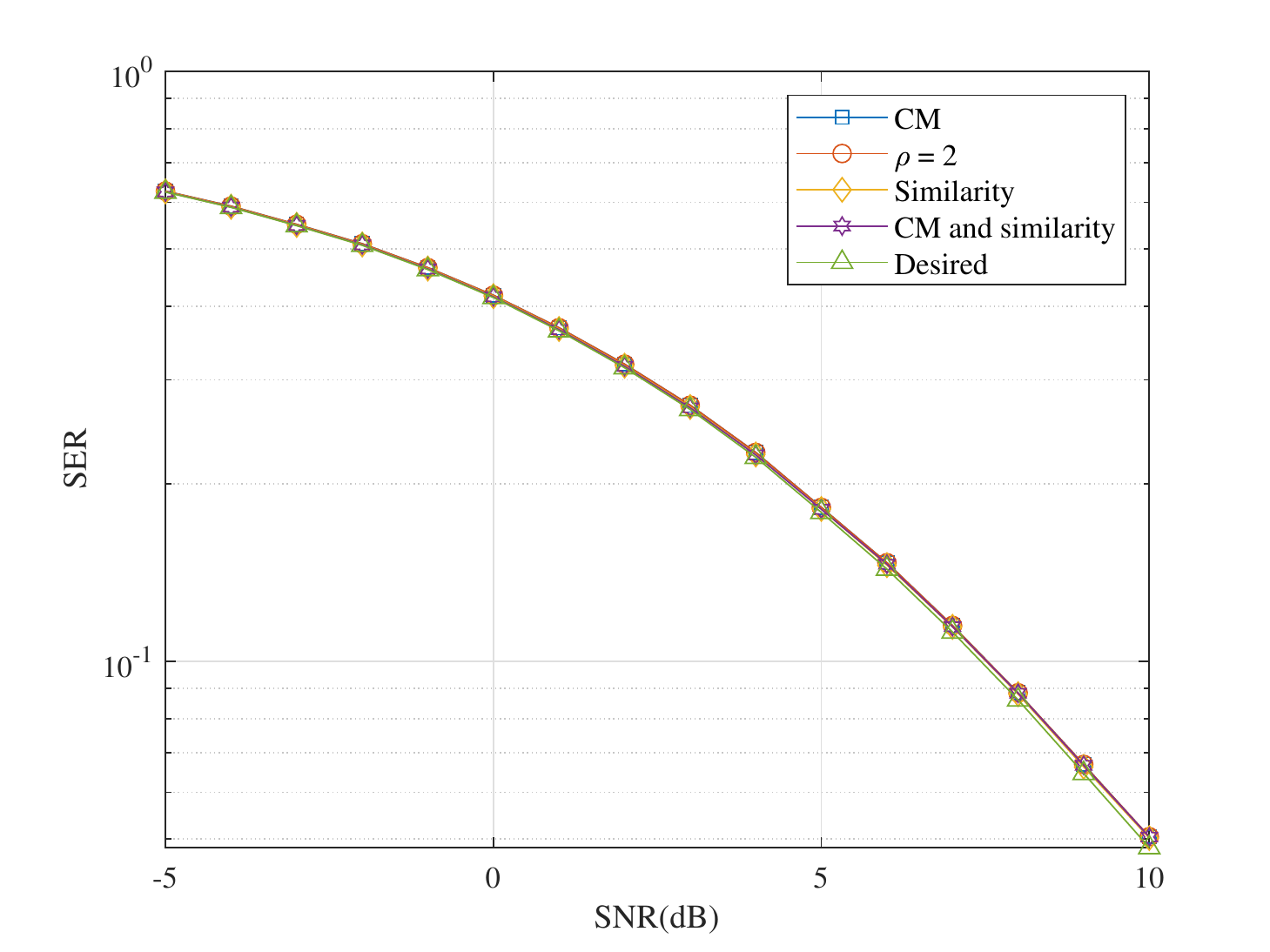}} \label{fig_14-2}} }
	\caption{SER under different constraints. $M=2, e_1 = e_2 = 20, \varsigma_1 = 10^{-3}, \varsigma_2 = 5\times 10^{-3}$. (a) User 1. (b) User 2.}
	\label{fig_14}
\end{figure*}

Finally, we demonstrate that the proposed algorithm can be extended to deal with different constraint (including the PAPR constraint, the similarity constraint, the CM and similarity constraints). We set $\rho = 2$, the similarity parameters $\delta = 0.25e_{\textrm{T}}$, and $\delta_{\infty} = 1.5\sqrt {p_s}$ \footnote{The similarity constraint is written as $\| \bx-\bx_0\|_2^2 \le \delta$; The constant-modulus and similarity constraints are written as $|\bx(n)| = \sqrt {p_s} , n=1, 2, \cdots,  LN_{\textrm{T}}, 	\| \bx-\bx_0\|_{\infty} \le \delta_{\infty}$.}. The reference waveform $\bx_0$ is given by:
\begin{equation}\label{eq:LFM}
	\bx_0(n) = \sqrt{p_s}\textrm{e}^{\textrm{j}\pi (n-1)^2/{LN_{\textrm{T}}}}, n=1, 2, \cdots,  LN_{\textrm{T}}.
\end{equation}
The other parameter setting is the same as that in \figurename~\ref{fig_2}. In \figurename~\ref{fig_13}, we can observe that increasing the PAPR results in a better detection performance. On the contrary, enforcing a similarity constraint will degrade the detection performance. \figurename~\ref{fig_14} analyzes the performance of the communication signals synthesized under different constraints. We can find that since all the synthesized communication signals satisfy the communication constraint, they have similar performance.

\section{Conclusion}\label{Sec:Conclusion}
An algorithm for jointly designing the transmit waveforms and receive filters was devised to maximize the detection performance of the DFRC system and ensure communications with multiple users.
Results showed that the optimized waveforms and filters could form deep nulls at the interference directions while maintaining the target response. In addition, the synthesized communication signals approximated the desired ones with the synthesis error of every user being precisely controlled. Moreover, if the transmit energy of the desired communication signals or the number of communication users was increased, the target detection performance degraded. Therefore, there is a fundamental tradeoff between the target detection and the communication performance.

\bibliographystyle{IEEEtran}

\bibliography{reference}

\begin{thebibliography}{10}
\providecommand{\url}[1]{#1}
\csname url@samestyle\endcsname
\providecommand{\newblock}{\relax}
\providecommand{\bibinfo}[2]{#2}
\providecommand{\BIBentrySTDinterwordspacing}{\spaceskip=0pt\relax}
\providecommand{\BIBentryALTinterwordstretchfactor}{4}
\providecommand{\BIBentryALTinterwordspacing}{\spaceskip=\fontdimen2\font plus
\BIBentryALTinterwordstretchfactor\fontdimen3\font minus
  \fontdimen4\font\relax}
\providecommand{\BIBforeignlanguage}[2]{{%
\expandafter\ifx\csname l@#1\endcsname\relax
\typeout{** WARNING: IEEEtran.bst: No hyphenation pattern has been}%
\typeout{** loaded for the language `#1'. Using the pattern for}%
\typeout{** the default language instead.}%
\else
\language=\csname l@#1\endcsname
\fi
#2}}
\providecommand{\BIBdecl}{\relax}
\BIBdecl

\bibitem{Griffiths2014spectrum}
H.~Griffiths, ``The challenge of spectrum engineering,'' in \emph{2014 11th
  European Radar Conference}, 2014.

\bibitem{Griffiths2015spectrum}
H.~Griffiths, L.~Cohen, S.~Watts, E.~Mokole, C.~Baker, M.~Wicks, and S.~Blunt,
  ``Radar spectrum engineering and management: technical and regulatory
  issues,'' \emph{Proceedings of the IEEE}, vol. 103, no.~1, pp. 85--102, 2015.

\bibitem{Aubry2016Optimization}
A.~Aubry, V.~Carotenuto, A.~De~Maio, A.~Farina, and L.~Pallotta, ``Optimization
  theory-based radar waveform design for spectrally dense environments,''
  \emph{IEEE Aerospace and Electronic Systems Magazine}, vol.~31, no.~12, pp.
  14--25, 2016.

\bibitem{Qian2018Coexistence}
J.~Qian, M.~Lops, L.~Zheng, X.~Wang, and Z.~He, ``Joint system design for
  coexistence of {MIMO} radar and {MIMO} communication,'' \emph{IEEE
  Transactions on Signal Processing}, vol.~66, no.~13, pp. 3504--3519, 2018.

\bibitem{Zheng2019Coexistence}
L.~Zheng, M.~Lops, Y.~C. Eldar, and X.~Wang, ``Radar and communication
  coexistence: An overview,'' \emph{IEEE Signal Processing Magazine}, vol.~36,
  no.~5, pp. 85--99, 2019.

\bibitem{howland2005passive}
P.~Howland, ``Passive radar systems,'' \emph{IEE Proceedings-Radar, Sonar and
  Navigation}, vol. 152, no.~3, pp. 105--106, 2005.

\bibitem{kuschel2019tutorial}
H.~Kuschel, D.~Cristallini, and K.~E. Olsen, ``Tutorial: Passive radar
  tutorial,'' \emph{IEEE Aerospace and Electronic Systems Magazine}, vol.~34,
  no.~2, pp. 2--19, 2019.

\bibitem{Jakabosky2016spectrum}
J.~Jakabosky, B.~Ravenscroft, S.~D. Blunt, and A.~Martone, ``Gapped spectrum
  shaping for tandem-hopped radar/communications \& cognitive sensing,'' in
  \emph{IEEE Radar Conference (RadarConf)}.\hskip 1em plus 0.5em minus
  0.4em\relax IEEE, 2016, Conference Proceedings, pp. 1--6.

\bibitem{Aubry2021Cognitive}
A.~Aubry, V.~Carotenuto, A.~D. Maio, M.~A. Govoni, and A.~Farina,
  ``Experimental analysis of block-sparsity-based spectrum sensing techniques
  for cognitive radar,'' \emph{IEEE Transactions on Aerospace and Electronic
  Systems}, vol.~57, no.~1, pp. 355--370, 2021.

\bibitem{Shi2022TAES}
C.~Shi, Y.~Wang, S.~Salous, J.~Zhou, and J.~Yan, ``Joint transmit resource
  management and waveform selection strategy for target tracking in distributed
  phased array radar network,'' \emph{IEEE Transactions on Aerospace and
  Electronic Systems}, vol.~58, no.~4, pp. 2762--2778, 2022.

\bibitem{Lindenfeld2004Sparse}
M.~J. Lindenfeld, ``Sparse frequency transmit-and-receive waveform design,''
  \emph{IEEE Transactions on Aerospace and Electronic Systems}, vol.~40, no.~3,
  pp. 851--861, 2004.

\bibitem{Rowe2014SHAPE}
W.~Rowe, P.~Stoica, and J.~Li, ``Spectrally constrained waveform design,''
  \emph{IEEE Signal Processing Magazine}, vol.~31, no.~3, pp. 157--162, 2014.

\bibitem{Aubry2014spectrally}
A.~Aubry, A.~De~Maio, M.~Piezzo, and A.~Farina, ``Radar waveform design in a
  spectrally crowded environment via nonconvex quadratic optimization,''
  \emph{IEEE Transactions on Aerospace and Electronic Systems}, vol.~50, no.~2,
  pp. 1138--1152, 2014.

\bibitem{Liang2015LPNN}
J.~Liang, H.~C. So, C.~S. Leung, J.~Li, and A.~Farina, ``Waveform design with
  unit modulus and spectral shape constraints via {Lagrange} programming neural
  network,'' \emph{IEEE Journal of Selected Topics in Signal Processing},
  vol.~9, no.~8, pp. 1377--1386, 2015.

\bibitem{Tang2019Efficient}
B.~Tang and J.~Liang, ``Efficient algorithms for synthesizing probing waveforms
  with desired spectral shapes,'' \emph{IEEE Transactions on Aerospace and
  Electronic Systems}, vol.~55, no.~3, pp. 1174--1189, 2019.

\bibitem{Li2016codesign}
B.~Li, A.~P. Petropulu, and W.~Trappe, ``Optimum co-design for spectrum sharing
  between matrix completion based {MIMO} radars and a {MIMO} communication
  system,'' \emph{IEEE Transactions on Signal Processing}, vol.~64, no.~17, pp.
  4562--4575, 2016.

\bibitem{Li2017Coexistence}
B.~Li and A.~P. Petropulu, ``Joint transmit designs for coexistence of {MIMO}
  wireless communications and sparse sensing radars in clutter,'' \emph{IEEE
  Transactions on Aerospace and Electronic Systems}, vol.~53, no.~6, pp.
  2846--2864, 2017.

\bibitem{Tavik2005RF}
G.~C. Tavik and I.~D. Olin, ``The advanced multifunction {RF} concept,''
  \emph{IEEE Transactions on Microwave Theory \& Techniques}, vol.~53, no.~3,
  pp. 1009--1020, 2005.

\bibitem{Zhang2022Survey}
J.~A. Zhang, M.~L. Rahman, K.~Wu, X.~Huang, Y.~J. Guo, S.~Chen, and J.~Yuan,
  ``Enabling joint communication and radar sensing in mobile networks-a
  survey,'' \emph{IEEE Communications Surveys \& Tutorials}, vol.~24, no.~1,
  pp. 306--345, 2022.

\bibitem{Zhang2021Overview}
J.~A. Zhang, F.~Liu, C.~Masouros, R.~W. Heath, Z.~Feng, L.~Zheng, and
  A.~Petropulu, ``An overview of signal processing techniques for joint
  communication and radar sensing,'' \emph{IEEE Journal of Selected Topics in
  Signal Processing}, vol.~15, no.~6, pp. 1295--1315, 2021.

\bibitem{Liu2020overview}
F.~Liu, C.~Masouros, A.~P. Petropulu, H.~Griffiths, and L.~Hanzo, ``Joint radar
  and communication design: Applications, state-of-the-art, and the road
  ahead.'' \emph{IEEE Transactions on Communications}, vol.~68, pp. 3834 --
  3862, 2020.

\bibitem{Shi2021DRFC}
C.~Shi, Y.~Wang, F.~Wang, S.~Salous, and J.~Zhou, ``Joint optimization scheme
  for subcarrier selection and power allocation in multicarrier dual-function
  radar-communication system,'' \emph{IEEE Systems Journal}, vol.~15, no.~1,
  pp. 947--958, 2021.

\bibitem{Tang2022MFRF}
B.~Tang and P.~Stoica, ``{MIMO} multifunction {RF} systems: Detection
  performance and waveform design,'' \emph{IEEE Transactions on Signal
  Processing}, vol.~70, pp. 4381--4394, 2022.

\bibitem{Hassanien2016DFRC}
A.~Hassanien, M.~G. Amin, Y.~D. Zhang, and F.~Ahmad, ``Dual-function
  radar-communications: Information embedding using sidelobe control and
  waveform diversity,'' \emph{IEEE Transactions on Signal Processing}, vol.~64,
  no.~8, pp. 2168--2181, 2016.

\bibitem{McCormick2017Simultaneous}
P.~M. McCormick, S.~D. Blunt, and J.~G. Metcalf, ``Simultaneous radar and
  communications emissions from a common aperture, part i: Theory,'' in
  \emph{IEEE Radar Conference (RadarConf)}, 2017, Conference Proceedings, pp.
  1685--1690.

\bibitem{Liu2018DFRC}
F.~Liu, L.~Zhou, C.~Masouros, A.~Li, and W.~Luo, ``Toward dual-functional
  radar-communication systems: Optimal waveform design,'' \emph{IEEE
  Transactions on Signal Processing}, vol.~66, no.~16, pp. 4264--4279, 2018.

\bibitem{Tang2020DFRC}
B.~Tang, H.~Wang, L.~Qin, and L.~Li, ``Waveform design for dual-function {MIMO}
  radar-communication systems,'' in \emph{2020 IEEE 11th Sensor Array and
  Multichannel Signal Processing Workshop (SAM)}, 2020, pp. 1--5.

\bibitem{Shi2020DFRC}
S.~Shi, Z.~Wang, Z.~He, and Z.~Cheng, ``Constrained waveform design for
  dual-functional {MIMO} radar-communication system,'' \emph{Signal
  Processing}, vol. 171, no.~1, p. 107530, 2020.

\bibitem{Liu2020Joint}
X.~Liu, T.~Huang, N.~Shlezinger, Y.~Liu, J.~Zhou, and Y.~C. Eldar, ``Joint
  transmit beamforming for multiuser {MIMO} communications and {MIMO} radar,''
  \emph{IEEE Transactions on Signal Processing}, vol.~68, pp. 3929--3944, 2020.

\bibitem{Tsinos2021DFRC}
C.~G. Tsinos, A.~Arora, S.~Chatzinotas, and B.~Ottersten, ``Joint transmit
  waveform and receive filter design for dual-function radar-communication
  systems,'' \emph{IEEE Journal of Selected Topics in Signal Processing},
  vol.~15, no.~6, pp. 1378--1392, 2021.

\bibitem{Larsson2013precoding}
S.~K. Mohammed and E.~G. Larsson, ``Per-antenna constant envelope precoding for
  large multi-user {MIMO} systems,'' \emph{IEEE Transactions on
  Communications}, vol.~61, no.~3, pp. 1059--1071, 2013.

\bibitem{Tang2020Polyphase}
B.~Tang, J.~Tuck, and P.~Stoica, ``Polyphase waveform design for {MIMO} radar
  space time adaptive processing,'' \emph{IEEE Transactions on Signal
  Processing}, vol.~68, pp. 2143--2154, 2020.

\bibitem{1967Dinkelbach}
W.~Dinkelbach, ``On nonlinear fractional programming.'' \emph{Management
  Science}, vol.~13, no.~7, pp. 492--498, 1967.

\bibitem{Boyd2010ADAMM}
S.~Boyd, N.~Parikh, E.~C. hu, B.~Peleato, and J.~Ec~Kstein, ``Distributed
  optimization and statistical learning via the alternating direction method of
  multipliers,'' \emph{Foundations \& Trends in Machine Learning}, vol.~3,
  no.~1, pp. 1--122, 2010.

\bibitem{Soltanalian2013Joint}
M.~Soltanalian, B.~Tang, J.~Li, and P.~Stoica, ``Joint design of the receive
  filter and transmit sequence for active sensing,'' \emph{IEEE Signal
  Processing Letters}, vol.~20, no.~5, pp. 423--426, 2013.

\bibitem{Soltanalian2014Optimization}
M.~Soltanalian and P.~Stoica, ``Designing unimodular codes via quadratic
  optimization,'' \emph{IEEE Transactions on Signal Processing}, vol.~62,
  no.~5, pp. 1221--1234, 2014.

\bibitem{Cui2017Quadratic}
G.~Cui, X.~Yu, G.~Foglia, Y.~Huang, and J.~Li, ``Quadratic optimization with
  similarity constraint for unimodular sequence synthesis,'' \emph{IEEE
  Transactions on Signal Processing}, vol.~65, no.~18, pp. 4756--4769, 2017.

\bibitem{Tsinos2022CM}
A.~Arora, C.~G. Tsinos, M.~R.~B. Shankar, S.~Chatzinotas, and B.~Ottersten,
  ``Efficient algorithms for constant-modulus analog beamforming,'' \emph{IEEE
  Transactions on Signal Processing}, vol.~70, pp. 756--771, 2022.

\bibitem{Tranter2017CM}
J.~Tranter, N.~D. Sidiropoulos, X.~Fu, and A.~Swami, ``Fast unit-modulus least
  squares with applications in beamforming,'' \emph{IEEE Transactions on Signal
  Processing}, vol.~65, no.~11, pp. 2875--2887, 2017.

\bibitem{Tang2021Information}
B.~Tang and P.~Stoica, ``Information-theoretic waveform design for {MIMO} radar
  detection in range-spread clutter,'' \emph{Signal Processing}, vol. 182,
  no.~5, p. 107961, 2021.

\bibitem{Tang2021Profiling}
B.~Tang, J.~Liu, H.~Wang, and Y.~Hu, ``Constrained radar waveform design for
  range profiling,'' \emph{IEEE Transactions on Signal Processing}, vol.~69,
  no.~8, pp. 1924--1937, 2021.

\bibitem{Palomar2018CM}
L.~Wu, P.~Babu, and D.~P. Palomar, ``Transmit waveform/receive filter design
  for {MIMO} radar with multiple waveform constraints,'' \emph{IEEE
  Transactions on Signal Processing}, vol.~66, no.~6, pp. 1526--1540, 2018.

\bibitem{Hong2016Convergence}
M.~Hong, Z.-Q. Luo, and M.~Razaviyayn, ``Convergence analysis of alternating
  direction method of multipliers for a family of nonconvex problems,''
  \emph{SIAM Journal on Optimization}, vol.~26, no.~1, pp. 337--364, 2016.

\bibitem{Li2006SPL}
J.~Li, J.~R. Guerci, and L.~Xu, ``Signal waveform's optimal-under-restriction
  design for active sensing,'' \emph{IEEE Signal Processing Letters}, vol.~13,
  no.~9, pp. 565--568, 2006.

\bibitem{Maio2008TSP}
A.~D. Maio, S.~D. Nicola, Y.~Huang, S.~Zhang, and A.~Farina, ``Code design to
  optimize radar detection performance under accuracy and similarity
  constraints,'' \emph{IEEE Transactions on Signal Processing}, vol.~56,
  no.~11, pp. 5618--5629, 2008.

\bibitem{Tang2016TxRx}
B.~Tang and J.~Tang, ``Joint design of transmit waveforms and receive filters
  for {MIMO} radar space-time adaptive processing,'' \emph{IEEE Transactions on
  Signal Processing}, vol.~64, no.~18, pp. 4707--4722, 2016.

\bibitem{kaybook1998}
S.~M. Kay, \emph{Fundamentals of Statistical Signal Processing-Volume {II}:
  Detection Theory}.\hskip 1em plus 0.5em minus 0.4em\relax New Jersey:
  Prentice Hall, 1998.

\bibitem{Pd1973}
L.~E. Brennan and I.~S. Reed, ``Theory of adaptive radar,'' \emph{IEEE
  Transactions on Aerospace and Electronic Systems}, pp. 237--252, 1973.

\end{thebibliography}

\newpage

\vfill

\end{document}